\documentclass[fleqn,useAMS, usenatbib]{mnras}

\usepackage[T1]{fontenc}
\usepackage{ae,aecompl}

 \usepackage{graphicx}	% Including figure files
 \usepackage{amsmath}	% Advanced maths commands
 \usepackage{amssymb}	% Extra maths symbols
 \usepackage{verbatim}
 \usepackage{textcomp}

  \usepackage{natbib}

 \newcommand{\be}{\begin{equation}}
 \newcommand{\ee}{\end{equation}}
 \newcommand{\ba}{\begin{eqnarray}}
 \newcommand{\ea}{\end{eqnarray}}
 \newcommand{\bs}{\begin{subequations}}
 \newcommand{\es}{\end{subequations}}

\title[Nutation Damping of Precessing Rotators]{Precession Relaxation of Viscoelastic Oblate Rotators}

\author[Frouard \& Efroimsky]
{Julien Frouard \thanks{Contact e-mail: \href{mailto:julien.frouard.ctr@navy.mil}{julien.frouard.ctr@navy.mil}}
~and~
Michael Efroimsky \thanks{Contact e-mail: \href{mailto:michael.efroimsky@navy.mil}{michael.efroimsky@navy.mil}}\\
US Naval Observatory, 3450 Massachusetts Ave NW, Washington DC 20392 USA \\
}

\date{Accepted XXX. Received YYY; in original form ZZZ}

\pubyear{2016}

\begin{document}
\label{firstpage}
\pagerange{\pageref{firstpage}--\pageref{lastpage}}
\maketitle

% Abstract of the paper
\begin{abstract}

 Perturbations of all sorts destabilise the rotation of a small body and leave it in a non-principal spin state. In such a state, the body experiences alternating stresses generated by the inertial forces. This yields nutation relaxation, i.e., evolution of the spin towards the principal rotation about the maximal-inertia axis. Knowledge of the timescales needed to damp the nutation is crucial in studies of small bodies' dynamics. In the literature hitherto, nutation relaxation has always been described with aid of an empirical quality factor $\,Q\,$ introduced to parameterise the energy dissipation rate. 

Among the drawbacks of this approach was its inability to describe the dependence of the relaxation rate upon the current nutation angle. This inability stemmed from our lack of knowledge of the quality factor's dependence on the forcing frequency. In this article, we derive our description of nutation damping directly from the rheological law obeyed by the material. This renders us the nutation damping rate as a function of the current nutation angle, as well as of the shape and the rheological parameters of the body. In contradistinction from the approach based on an empirical $\,Q\,$-factor, our development gives a zero damping rate in the spherical-shape limit. Our method is generic and applicable to any shape and to any linear rheological law. However, to simplify the developments, here we consider a dynamically oblate rotator with a Maxwell rheology.

\end{abstract}

% Select between one and six entries from the list of approved keywords.
% Don't make up new ones.

%%%%%%%%%%%%%%%%%%%%%%%%%%%%%%%%%%%%%%%%%%%%%%%%%%
%%%%%%%%%%%%%%%%% BODY OF PAPER %%%%%%%%%%%%%%%%%%

\section{Motivation}

 \cite{prendergast1958} and, later, \cite{burns1971} drew attention to the fact that alternating dissipative forces emerging in a precessing rotator must reduce its kinetic energy without affecting its angular momentum. With no external torques acting on the body, its nonprincipal rotation must relax to motion about the maximal-inertial axis~---~a spin state corresponding to the minimal energy with a fixed angular momentum. The then available statistics on asteroids contained only singly-periodic curves, warranting that all the observed asteroids were in the final spin state.

 In the turbulent antecedents of the solar system, there were {\ae}ons when collisions and disruptions were regular, so asteroids (and their smithereens) were set into wobble often. Even in the present epoch, there happen events capable of kicking asteroids out of principal spin. Such events include occasional collisions, as well as tidal interactions during asteroids' close flybys near planets. Small bodies can be driven into nonprincipal axis (NPA) spin also by the YORP effect~---~see, e.g., \citet{vokrou+2007} and Breiter, Ro{{\.{z}}}ek \& Vokrouhlick{\'{y}} (\citeyear{breiter+2011}) and references therein. Wobble of comets is impelled mainly by jetting. Gradual outgassing, too, may contribute to the effect, because a rotator goes into tumbling when it changes its principal axes through a partial loss or redistribution of the material.

In an excited state, an unsupported top tends to damp the excessive energy, while keeping the angular momentum unchanged.  Energy damping is caused by the alternating stresses emerging in an excited rotator due to the transversal and centripetal acceleration of its parts. The stresses periodically deform the body, and internal friction dissipates energy. This process is called $\,${\it{inelastic dissipation}}. It yields evolution of the spin state towards the principal rotation (the one about the maximal-inertia axis).$\,$\footnote{~Besides the inelastic dissipation, there also exists a so-called $\,${\it{Barnett dissipation}}$\,$ caused by periodic remagnetisation of a precessing top. The mechanism is important only for cosmic dust granules \citep{lazarian1997}, and we shall not discuss it here.}

 As of now, 144 out of 16444 entries in the LCDB light curve database Warner, Harris \& Pravec (\citeyear[the April 2017 version]{warner+2009}) contain a signature of NPA rotation, while the actual percentage of tumblers is expected to be even larger.$\,$\footnote{~As detection of NPA rotation is very observationally demanding, there is a strong observational bias against them in the LCDB database. Accounting for the bias, \citet{pravec2014} argue that most of the asteroids, whose size is larger than about $\,0.2$ km and whose characteristic damping timescale is longer than about $\,0.2$ Gyr, are actually in NPA spin states. While most of the large tumblers are spinning slowly, the situation is different for asteroids smaller than $\,0.2$ km. In the case of such small sizes, NPA rotations are frequent also among very fast spinners.  (Petr Pravec, private communication.)
 See, e.g., \citet{pravec2005} and \citet{scheirich2010} for examples of two small super-fast spinning tumblers, 2000 WL107 and 2008 TC$_3$.
 } The question arises when these rotators acquired their non-principal spin. Could any of them have preserved their primordial wobble?

 Even more tempting would be to try to infer from the collected statistics some useful information on the physical properties of asteroids, such as viscosity and/or rigidity.

 \section{History}\label{hi}

 \cite{burnssafronov1973} enquired how long it would take a ``typical" asteroid to relax from wobble to the principal spin. Using back-of-the-envelope estimates, they obtained a formula for a typical timescale of relaxation:
 \begin{equation}
 \tau \propto \frac{\mu~Q}{\rho\,R^2\,\Omega^3}~A~,
 \label{1}
 \end{equation}
 with $\mu, Q, \rho, R$ being the mean shear rigidity, quality factor, density and radius of the body; and with $\Omega$ being the spin rate. The numerical factor was estimated to be about a hundred
 % (see {\it{Ibid.}}, eqns 22 - 23):
 \citep[eqns. 22 - 23]{burnssafronov1973}:
\begin{equation}
A^{ \rm  \textstyle{^{ \rm  (B\&S)}}} \propto 100~.
\label{2}
\end{equation}
Using these formulae, the authors arrived at a conclusion that, for most asteroids, the relaxation times are much shorter than the age of the solar system, though in some cases the relaxation times could be comparable to that age.

A more accurate study was later performed by \cite{efroimskylazarian2000} for oblate rotators. The authors wrote down the wobble-generated inertial force in an arbitrary point of the body, then found a distribution of stresses corresponding to the obtained distribution of force. After that, they derived the ensuing strain and, finally, obtained the dissipation rate, using the empirical quality factor for solids. As was pointed out in \cite{efroimskylazarian2000}, a large part of dissipation comes from the second harmonic which is a double of the precession frequency.$\,$\footnote{~The second harmonic shows up as the centrifugal force is quadratic in the angular velocity $\,\mathbf{\Omega}\,$. For an oblate body, the components of $\,\mathbf{\Omega}\,$ are proportional to $\,\sin \omega t\,$ and $\,\cos\omega t\,$, where $\,\omega\,$ is the precession rate. Hence the squaring of $\,\mathbf{\Omega}\,$ furnishes terms with the sine and cosine of $\,2 \omega t\,$.} That analysis rendered a much faster relaxation rate:
\begin{equation}
A^{ \rm  \textstyle{^{ \rm  (E\&L)}}} \propto 1\;\;\mbox{---}\;\;4 .
\label{3}
\end{equation}
In \cite{efroimsky2000,efroimsky2001,efroimsky2002}, a similar approach was used for triaxial bodies.

Although a step forward, the analysis in \cite{efroimskylazarian2000} and \cite{efroimsky2000,efroimsky2001,efroimsky2002} still was not fully rigorous. First, the rotator was modeled with a rectangular prism, a shape too dissimilar from that of actual asteroids. Second, the boundary conditions for the stresses were satisfied on the surfaces only approximately.

The first of these two deficiencies was soon fixed by Molina, Moreno \& Mart\'{i}nez-L\'{o}pez (\citeyear{molina+2003}) who applied the above method to an oblate ellipsoid (still keeping the boundary conditions approximate). They arrived at a range of values somewhat in between the results obtained by \cite{efroimskylazarian2000} and \cite{burnssafronov1973}:
\begin{equation}
A^{ \rm  \textstyle{^{ \rm  (Molina)}}} \propto 10 \;\;\mbox{---}\;\;30~.
\label{4}
\end{equation}

 A more elaborate treatment was then suggested by Sharma, Burns \& Hui (\citeyear{sharma+2005}) who solved the equation for displacements. These authors got, for oblate rotators:
 \begin{equation}
 A^{ \rm  \textstyle{^{ \rm  (Sharma)}}} \propto 200 \;\;\mbox{---}\;\; 800~,
 \label{5}
 \end{equation}
 and even larger values for prolate bodies.

 Later research by Breiter, Ro{{\.{z}}}ek \& Vokrouhlick{\'{y}} (\citeyear{breiter+2012}) demonstrated that the inflated values of $\,A\,$ were obtained by \cite{sharma+2005} due to accumulation of three oversights\footnote{According to
 \cite{breiter+2012}, a numerical oversight in \cite{sharma+2005} resulted in a redundant factor of $\pi$, while an unconventional definition of $Q$ caused a factor of 2. So $\,A^{ \rm  \textstyle{^{ \rm  (Sharma~et~al)}}}$ should have been divided by $2\pi$, and the ensuing relaxation time would no longer be unusually high. It would still remain higher than the right answer, because \cite{sharma+2005} did not account for an extra multiplier of 2 in the work by 2nd harmonic. This is illustrated by Figures 5 and 6 in \cite{breiter+2012}. There, the solid curves correspond to the exact solution obtained in \cite{breiter+2012}, while the dot-dashed curves depict the solution from \cite{sharma+2005}.}. \cite{breiter+2012} also showed that the shorter damping times obtained by \cite{efroimskylazarian2000} resulted mainly (factor $\,14/\pi\,$) from modeling the rotator with a prism of a volume higher than any solid of revolution with the same ratio of axes.
 
 In all those publications, the quest was for new skill in the analytic or computational treatment of the thitherto existing formulation of the problem, rather than for added physical insight into the subject. Specifically, in each of those works, calculation comprised two steps. First, it was taken for granted that the boundary value problem for stresses and strains (or displacements) can be solved under the assumption that the rotator is elastic~---~so the instantaneous strain tensor is linear in the instantaneous stress. Then, after the stresses and strains (or displacements) were calculated, an empirical quality factor $\,Q\,$ was introduced, to account for the energy damping rate (with the intention to derive, subsequently, the corresponding decay of the precession cone).

 These two steps were incompatible. Indeed, elasticity implies instantaneous reaction, i.e., vanishing of the phase lag between the strain and stress. At the same time, calculation of the power dissipated at a certain frequency shows that the inverse quality factor is equal to the sine of the phase lag at this frequency~---~see Appendix \ref{AppendixA} below.

 This inherently contradictive approach may, arguably, be applied to near-elastic media. There, the rigidity overwhelmingly defines the deformation, and the quality factor serves to empirically amend for the neglect of a small viscosity in that calculation. This method, however, is inapplicable to asteroids and other small bodies. In treatment of those objects, the effective viscosity is an important parameter which should from the beginning enter the calculation of deformation caused by the stress \citep{efroimsky2015}. This calculation should furnish the phase lag at each frequency involved. The knowledge of these lags will then render the energy dissipation rate at each mode.

 In this paper, we abandon the idea of introducing the {\it{ad hoc}} quality factor $\,Q\,$. Instead, we use a solution to the boundary value problem for stresses and strains in a {\underline{{visco}}}elastic rotator, in order to obtain the dissipated power (and, thence, the rate of nutation damping).
 This approach enables us, among other things, to express the relaxation rate and the relaxation timescale as functions of the residual nutation angle.
 In contradistinction from the preceding works, we obtain a finite dissipation rate in the limit of a spherical shape.

 In our study, we accept the viscoelastic Maxwell model, though our formalism can be combined with an arbitrary linear rheology. We also assume that the unperturbed (no-wobble) shape of the body is oblate~---~which is an equilibrium shape of a viscoelastic rotator. Our formalism can be extended to a triaxial equilibrium shape,$\,$\footnote{~For low rotation rate, the figure of (stable) equilibrium of a mass of inviscid fluid is oblate (and is known as a $\,${\it{Maclaurin ellipsoid}}). However, once a specific rotation rate is attained, a triaxial shape is possible. It is named $\,${\it{Jacobi
 ellipsoid}}$\,$ and is stable \citep{grig}. Other shapes at higher rotation rates are possible but unstable.
 \label{footnote4}} by using the stress tensor from \citet{breiter+2012}.

 Rigorously speaking, ellipsoid is no longer a figure of equilibrium when the spin axis goes far away from the maximum-inertia axis. We, however, assume that, once formed in a principal state, the body is capable to sustain its shape (aside from small periodic deformations). So we use the Maxwell model (or whatever other linear rheology) for small deformations only, assuming that at large deformations the reaction is more persistent.

 \section{Outline of the method}

 To compute the damping time of a nutating rotator, we need to know the distribution of stress in this body, with both the inertial forces and self-gravitation taken into account.
 We shall be interested in two cases~---~that of an oblate spheroid \citep{sharma+2005} and that of a rectangular prism \citep{efroimskylazarian2000}. In this section, we discuss how the boundary value problem for stress was approached in these works and, specifically, what underlying assumptions were adopted by those authors.

\subsection{Precession of an unsupported top.\\ Deviation from rigidity}\label{precession}

 We consider the free motion of a top, as seen in the body frame. Our coordinate system is defined by the three principal axes of inertia: $\,1,2,3\,$; with coordinates $\,x,y,z\,$; and with unit vectors $\,\mathbf{e}_{ \rm  1},\,\mathbf{e}_{ \rm  2},\,\mathbf{e}_{ \rm  3}\,$. The angular velocity is denoted by $\,\mathbf{\Omega}\,$, while $\,\omega\,$ signifies the rate of precession. We begin with the Euler equations
 \begin{equation}
 \frac{d}{dt} \left(I_i  \Omega_i \protect\right)  ~=~  \left(I_j  - I_k \protect\right) \Omega_j  \Omega_k\;,
 %\label{2.1}
 \label{6}
 \end{equation}
 where $\,(i, j, k)=(1, 2, 3)\,$, with cyclic transpositions. The moments of inertia obey $\,I_1 \leq  I_2 \leq  I_3\,$. In the solid-body approximation, we set $\,\dot{I}_i  \Omega_i \ll I_i {\dot{\Omega}}_i\,$, and the equations simplify to
 \begin{equation}
 I_i{\dot{\Omega}}_i = \left(I_j - I_k \protect\right) \Omega_j \Omega_k\;.
 %\label{2.3}
 \label{8}
 \end{equation}
 The neglect of ${\dot{I}}_i \Omega_i$ against $I_i {\dot{\Omega}}_i$, however, comes into contradiction with our intention to bring internal dissipation into the picture. Dissipation results from small periodic deformations called into being by precession-generated stresses. These deformations yield periodic changes in the principal moments of inertia, as well as oscillations of the principal inertia axes; hence nonzero $\,\dot{I_i}
 \;$. \footnote{~We do not mention here of the emergence of nondiagonal terms in the inertia tensor, because they are very small and do not change our order-of-magnitude estimate. We could as well have justified our neglect of the nondiagonal terms by agreeing to stay in the principal-axes frame (whose orientation slightly oscillates relative to a body of an oscillating shape). The slight trembling of the principal-axes frame will produce small inertial torques on the right-hand side of the Euler equations, but will not alter our conclusions about the form of the left-hand side thereof.} To prove that $\dot{I_i}/{I_i} \ll \dot{\Omega_i}/{\Omega_i}$, recall that
 \begin{equation}
 {\dot{\Omega}}_i/{\Omega}_i  \approx  {\tau}^{ \rm  -1} \; , \; \; \; \; \;
 {\dot{I}}_i/{I}_i \approx  {\tau}^{ \rm  -1}  \epsilon \; \,,
 \label{9}
 \end{equation}
 with $\tau$ being a rotational period, and $\epsilon$ being a typical value of strain. Linear deformation usually takes place within the realm of $\epsilon < 10^{ \rm  -6}$. Therefore, insofar as the linear description is applicable, it is also legitimate to assume that
 \begin{equation}
 \dot{I}_i  \Omega_i \ll I_i  {\dot{\Omega}}_i\;\,,
 %\label{condition1}
 \label{10}
 \end{equation}
 the approximation (\ref{8}) thus being justified.

 A solution to the equations (\ref{8}) is parameterised by the values of the integrals of motion, which are the angular momentum $\,\bf{J}\,$ and the kinetic energy $\,T_{ \rm  kin}\,$:
 \begin{eqnarray}
 I_1^2  {\Omega}_1^2  + I_2^2  {\Omega}_2^2  +  I_3^2  {\Omega}_3^2  =  {\bf{J}}^2 \;, \;\;
 \label{11}\\
 I_1  {\Omega}_1^2  + I_2 {\Omega}_2^2  +  I_3  {\Omega}_3^2  =  2  T_{ \rm  kin} \;.
 \label{12}
 \end{eqnarray}
 In the space of the body-frame angular velocities $\,\Omega_i\,$, the equation (\ref{11}) parameterises an ellipsoid of angular momentum, its size being defined by the value of $\,{\bf{J}}^2\,$. A particular solution $\,\Omega_i(t)\,$ to the Euler equations (\ref{8}) will coincide with a closed line made by intersection of the angular-momentum ellipsoid with the ellipsoids of kinetic energy, parameterised by the equation (\ref{12}). A fixed angular-momentum ellipsoid (the one corresponding to a fixed $\,{\bf{J}}^2\,$) will be intersected differently by kinetic-energy ellipsoids corresponding to different values of $\,T_{ \rm  \small{kin}}\,$.

 The kinetic-energy ellipsoid corresponding to the largest available value of $\,T_{ \rm  \small{kin}}\,$ (largest~---~for the fixed $\,{\bf{J}}^2\,$) embeds
 the angular-momentum ellipsoid, touching it from outside in the two points which correspond to rotation about the minimal-inertia axis. The kinetic-energy ellipsoid corresponding to the smallest available value of $\,T_{ \rm  \small{kin}}\,$ is located inside the angular-momentum ellipsoid, and is touching it from inside in two opposite points. These two points correspond the states of complete relaxation of precession, i.e., to rotation about the maximal-inertia axis.

  To describe relaxation, we permit for slow leakage of energy, with the angular momentum being fixed. Tuning one of the integrals of motion in the course of motion is like changing the rules in the middle of the game. To justify this, we have to ensure adiabaticity, i.e., to be certain that the leakage (relaxation) is much slower a process than the precession (and, thereby, than the rotation). For a dynamically oblate top ($I_1 = I_2$), the timescale separation condition reads as
  \ba
  -\;\left(\frac{d \theta}{dt} \right)_{ \rm  (oblate)}\;\ll\;\omega\;,
  \label{13}
  \ea
  $\omega\,$ being the precession rate and $\,\theta\,$ being the angle between the angular momentum and the major-inertia axis. The condition owes its simple form to the fact that for an oblate top the angle $\,\theta\,$ stays constant over a precession cycle; so no averaging is needed, unlike in the more involved case of a triaxial top.$\,$\footnote{~As was demonstrated in \cite{efroimsky2000}, for a triaxial rotator the timescale separation condition becomes
 \ba
 - \frac{d  \langle \sin^2 \theta \rangle}{dt} \ll \chi_1\;,
 \nonumber
 \ea
 where $\,\langle.\,.\,.\rangle\,$ signifies time averaging over a precession period. Averaging is needed, because in the triaxial case the angle $\,\theta\,$ is periodically changing in time. Also mind that in the triaxial case we denote the precession frequency with $\,\chi_1\,$. The reason for this change of notation is explained in the end of Section \ref{kr}.}

While the condition (\ref{10}) justifies the approximate form of the equations of motion, the condition (\ref{13}) validates our by-hand compensation of that approximation. We hope to find a right solution by introducing a compensation approximation (energy leaking) into a formalism that was made approximate by the neglect of deformation.

 Usually in the literature the fulfilment of both conditions is taken for granted.$\,$\footnote{~It is, for example, seldom noted that the Euler equations in the form of (\ref{8}) are approximate when applied to realistic (not fully rigid) bodies.} In reality, the applicability domains of the conditions (\ref{10}) and (\ref{13}) are defined by the properties of the medium. While both these conditions are usually obeyed by rotators composed of viscoelastic materials, it is not immediately apparent if these conditions are always satisfied by rubble piles~---~especially by those whose spin rate is close to that of disruption. Nor is it evident whether the two conditions are independent from one another. It may be interesting to explore if the condition (\ref{10}) entails (\ref{13}).

 \subsection{Parameterisation of an ellipsoid and of the forces acting on it}

 \subsubsection{Notation}\label{notation}

 We model an asteroid by a homogeneous ellipsoid of mass $m$, with semiaxes $a = b \geq c$. We also introduce the dimensionless parameter $h=c/a$, so the moments of inertia can be expressed as
 \begin{equation}
 I\;=\;I_1\,=\;I_2\,=\;\frac{m\,a^2}{5} \left ( 1\,+\,h^2 \right) \;\;, \quad I_3 = \frac{2\,m\,a^2}{5} \;\;.
 \label{23}
 \end{equation}
 and the Euler equations (\ref{8}) acquire the form
 \begin{equation}
 \dot{\bold{\Omega}} = - \frac{\textstyle 1 - h^2}{\textstyle 1 + h^2} \,\Omega_{ \rm  2}\, \Omega_{ \rm  3}\, \mathbf{e}_{ \rm  1}
 + \frac{\textstyle 1 - h^2}{\textstyle 1 + h^2}\, \Omega_{ \rm  1} \,\Omega_{ \rm  3}\, \mathbf{e}_{ \rm  2} \;\;.
 \label{24}
 \end{equation}
 Here the unit vectors $ \mathbf{e}_{ \rm  1}, \mathbf{e}_{ \rm  2}, \mathbf{e}_{ \rm  3}$ are pointing along the principal axes of inertia. The corresponding coordinates $x,y,z$ can be reparameterised via new variables $\,\theta,\,\phi,\,q\,$ as
\begin{eqnarray}
 x & = & q \, a \sin \theta \cos\phi\;\;, \nonumber \\
 y & = & q \, b \sin \theta \sin\phi \;\;,\label{25} \\
 z & = & q \, a \, h \cos \theta \;\; \nonumber ,
\end{eqnarray}
 where $0 \leq \theta \leq \pi$, $0 \leq \phi < 2 \pi$, $0 \leq q \leq 1$, and the boundary is attained at $q=1$. Then the position of a point inside or on the body can be written down as
 \begin{eqnarray}
 \nonumber
 \bold{r} & = & x ~\mathbf{e}_{ \rm  1} +  y~ \mathbf{e}_{ \rm  2} + z~ \mathbf{e}_{ \rm  3} \\
          & = & q~a \left(\sin \theta \cos \phi~ \mathbf{e}_{ \rm  1} + \sin \theta \sin\phi ~\mathbf{e}_{ \rm  2} + h \cos \theta ~\mathbf{e}_{ \rm  3} \right) \;\;.\quad
\label{26}
 \end{eqnarray}

  Vibratory deformation of the body leaves the centre of mass quiescent. It always serves as the origin, though the directions of the principal axes oscillate slightly. We neglect those oscillations, insofar as the strain $\epsilon$ entering the condition (\ref{10}) stays small enough.

 An element of volume transforms as $dV = dx\,dy\,dz = {\cal{J}}\,dq\,d \theta\,d\phi$, where the Jacobian (derived from the expressions \ref{25}) is equal to $\,{\cal{J}}\,=\,a^3\,q^2\,h\, \sin \theta\,$. Accordingly,
 \begin{equation}
 \int dV =  a^3 \, h \int_{ \rm  0}^{ \rm  1} q^2\,dq\;\int_{ \rm  0}^{ \rm  \pi}\sin\theta\;d\theta\,\int_{ \rm  0}^{ \rm  2\pi}d\phi\;\;.
 \label{27}
 \end{equation}

 \subsubsection{Forces acting in a rotating body}

 To find the dissipation rate, we should know the distribution of the strain and stress in the body. The strain can be found from the stress, with aid of a rheological equation for the material. The stress can be found from the distribution of the reaction forces that emerge in the material. To calculate the reaction force $\bold{f}$ acting per unit mass, we recall that this force, together with gravity force $\bold{b}_{ \rm  gr}$, endows a unit mass with the acceleration $\bold{a}$ relative to an inertial frame:
 \begin{equation}
 {\bf{a}}\;=\;{\bf{f}}\;+\;{\bf{b}}_{ \rm  gr}\;\;.
 \label{28a}
 \end{equation}
 This is the Second Law of Newton in an inertial frame.  In the frame comoving with an element of medium,$\,$\footnote{~In neglect of the body-frame-related acceleration $\,\bf{a}\,'\,$, the locally comoving frame can be identified with the body frame. The neglect of $\,\bf{a}\,'\,$ will be justified in  Section \ref{3.2.3}.} this law can be rewritten as
 \ba
 -~{\bf{f}}~=~{\bf{b}}_{ \rm  gr}~-~{\bf{a}}\;\;.
  \label{28b}
 \ea
 \label{28}
 where $\,{\bf{b}}_{ \rm  gr}-\,{\bf{a}}\,$ is the overall body force.

 \subsubsection{The acceleration relative to an inertial frame, expressed through the body-frame-related angular velocity\label{3.2.3}}

 Our first step is to calculate the acceleration experienced by a point of the body. Although we need the acceleration with respect to an inertial frame, it is practical to express it in the body frame defined by the principal axes $\,(1,2,3)\,$, with the unit vectors $\,\mathbf{e}_{ \rm  1},\,\mathbf{e}_{ \rm  2},\,\mathbf{e}_{ \rm  3}\,$.

 In the principal axes, the angular velocity reads
  as
 \ba
 \label{}
 \bold{\Omega}\;=\;\Omega_{ \rm  1}\,\mathbf{e}_{ \rm  1}\,+\;\Omega_2\,\mathbf{e}_{ \rm  2}\,+\;\Omega_3\,\mathbf{e}_{ \rm  3}~~.
 \ea
  The position, velocity and acceleration in the body frame are termed $\,{\bf{r}},\,{\bf{v}},\,{\bf{a}}\,'\,$. We endow the body-frame-related acceleration with a prime, to distinguish it from the acceleration in the inertial frame. The latter acceleration is denoted with $\,{\bf{a}}\;$:
 \begin{equation}
 \bold{a} = \bold{a}^{ \rm  \prime} + \bold{\dot{\Omega}} \times \bold{r} + 2 ~\bold{\Omega} \times \bold{v} +   \bold{\Omega}  \times ( \bold{\Omega} \times \bold{r}) \;\;.
 \label{29}
 \end{equation}
  In Section \ref{precession}, the smallness of a typical strain $\epsilon$ justified our neglect of $\dot{I_j}\,\Omega$ against ${I_j}\,{\dot{\Omega}}\,$ in the Euler equations. On similar grounds, we neglect the first and third terms on the right-hand side of the above expression. Rotation with a period $\tau$ of a body of size $\,\mathit{l}\,$ generates deformation $\,\delta{\mathit{l}}\,\approx\,\epsilon\,{\mathit{l}}\,$, deformation-caused velocity $\,v\,\approx\,\delta\mathit{l}/\tau\,\approx\,\epsilon\,\mathit{l}/\tau\,$, and deformation-caused acceleration $\,a\,'\,\approx\,\delta\mathit{l}/\tau^{ \rm  2}\,=\,\epsilon\, \mathit{l}/\tau^{ \rm  2}\,$. These typical values of $\,v\,$ and $\,a'\,$ are much smaller than the velocity and acceleration of the body as a whole (that are about $\,{\mathit{l}}/\tau\,$  and $\,{\mathit{l}}/{\tau}^2\,$, correspondingly). Then
 \begin{equation}
 \bold{a} \approx \bold{\dot{\Omega}} \times \bold{r} +  \bold{\Omega}  \times ( \bold{\Omega} \times \bold{r}) \;\;.
 \label{30}
 \end{equation}

 \subsubsection{Rotation of a dynamically oblate rotator.\\
 Acceleration of a point inside such a rotator}\label{kr}

 For a rotator possessing an oblate dynamical symmetry,
 \be
 I_3 >I_2 = I_1\;,
 \ee
 the equations of motion (\ref{24}) can be easily solved analytically. It turns out that, in the body frame, both the angular velocity vector $\,\bold{\Omega}\,$ and the angular momentum vector $\,\bold{J}\,$ perform a circular precession about the axis 3 (the one of of maximum inertia). The components of $\,\bold{\Omega}\,$ read as
 \ba
 \Omega_1\,=\,\pm\,\Omega_{ \rm  \perp}\,\cos\omega t~,~~\Omega_2\,=\,\Omega_{ \rm  \perp}\,\sin\omega t~,~~\Omega_3\,=\,\text{const}~,
 \label{eq}
 \ea
 where
 \begin{equation}
 \Omega_{ \rm  \perp}\,=\,\Omega\;\sin\alpha\quad,\quad \Omega_{ \rm  3}\,=\,\Omega\;\cos\alpha \;,
 \end{equation}
 $\alpha\,$ being the angle between the angular velocity vector $\bold{\Omega}$ and the maximal-inertia axis. The precession frequency $\,\omega\,$ is given by
 \begin{equation}
 \omega = (H-1) \Omega_3\;\;,\quad\mbox{with}\;\;H\;\equiv\;\frac{I_3}{I_1}\;.
 \label{omegafreq}
 \end{equation}
 For an oblate spheroid, the expressions (\ref{23}) yield:
 \ba
 H\;=\;\frac{2}{1\;+\;h^2}\;.
 \nonumber
 \label{H}
 \ea
 The angle $\,\theta\,$ between the angular momentum vector $\,\bold{J}\,$ (with the norm $J$ = |\textbf{J}|) and the maximal-inertia axis obeys
 \begin{equation}
 \Omega_3\,=~\frac{J}{I_3}\;\cos\theta\quad,\qquad\Omega_{ \rm  \perp}\,=~\frac{J}{I_3}~H\;\sin\theta\;.
 \label{omegafreq}
 \end{equation}

 Insertion of the expression (\ref{eq}) into the equation (\ref{30}) makes it evident that the inertial forces contain two harmonics only: $\,\chi_1\,=\,\omega\,$ and $\,\chi_2\,=\,2\,\omega\,$, where $\,\omega\,$ is the precession frequency as seen in the body frame. Naturally, the ensuing expressions for the stress and strain (presented in Appendix \ref{shapes}) include only these two harmonics. This observation, however, pertains to dynamically oblate rotators only. In the triaxial case, the expressions for $\,\Omega_i\,$ are different from (\ref{eq}) and are given by elliptic functions $\,$sn$\,$, $\,$cn$\,$ and $\,$dn$\,$. Expansion of those into Fourier series renders an infinite number of harmonics. Consequently, insertion of those expressions into the equation (\ref{30}) yields an infinite spectrum of forcing frequencies that are harmonics of the lowest frequency: $\,\chi_n=n\chi_1\,$ \citep{efroimsky2000}.

 \subsubsection{Self-gravitation}

 The gravity force  $\bold{b}_{ \rm  gr}$ in a point inside a precessing rotator comprises two parts:
  \begin{equation}
 {\bf{b}}_{ \rm  gr}\,=~\overline{\bf{b}}_{ \rm  gr}\,+~{\textbf{\~{b}}}_{ \rm  gr}\;\;.
 \label{31}
  \end{equation}
 The small oscillating part $\,{\textbf{\~{b}}}_{ \rm  gr}\,$ is generated by the periodically evolving deformation caused by the inertial forces. It is {\underline{not}} $\,${\it{a priori}}$\,$ obvious whether it can be neglected.

 The constant part $\,\overline{\bf{b}}_{ \rm  gr}\,$ is simply the gravity force of an undeformed ellipsoid. In a point $x,y,z$ inside a spheroid
 (and, more generally, inside a triaxial ellipsoid of dimensions $a \geq b \geq c$), it is linear in the body coordinates:
  \begin{equation}
 \overline{\bf{b}}_{ \rm  gr}\,=\;-\;\gamma_1\,x~\mathbf{e}_{ \rm  1}\,-\;\gamma_2\,y~\mathbf{e}_{ \rm  2}\,-\;\gamma_3\,z~\mathbf{e}_{ \rm  3}\;\;.
 \label{32}
  \end{equation}
 Derived by \cite{gauss1813} and \cite{rodrigues1816}, the coefficients $\,\gamma_i\,$ can be found, e.g., in \citet{breiter+2012}.

  \subsubsection{The reaction force}

 In neglect of the oscillating part $\,\tilde{\bf{b}}_{ \rm  gr}\,$ of the gravity force, equation (\ref{28b}) becomes
 \footnote{\cite{breiter+2012} employed the notation $\bold{b}_{ \rm  in} = - \bold{a}$ and used $\bold{b}$ for the overall body force (that is, negative the reaction $\bold{f}$). They also ignored the oscillating part of gravity, i.e., did not distinguish between $\bold{b}_{ \rm  gr}$ and $\overline{\bold{b}}_{ \rm  gr}$. In this notation and approximation, our formula (\ref{36}) becomes
 \begin{equation}
 \bold{b} = \bold{b}_{ \rm  gr} + \bold{b}_{ \rm  in} = \mathbb{B}~\bold{r} \;,
 \nonumber
 \end{equation}
 which coincides with the equation (26) in \cite{breiter+2012}.}
 \begin{equation}
 - \bold{f}~ = ~ \bold{b}_{ \rm  gr} - \bold{a}~ =~ (\tilde{\bf{b}}_{ \rm  gr} + \overline{\bf{b}}_{ \rm  gr} ) - \bold{a} ~\approx ~ \overline{\bf{b}}_{ \rm  gr}\,-~{\bf{a}}~=~\mathbb{B} ~ \bold{r} \;,
 \label{36}
 \end{equation}
 The elements of the matrix $\mathbb{B}$ are given by the expressions~\footnote{~Using the notations $\,h_1 \equiv b/a~$ and $~h_2 \equiv c/b\,$, \cite[eqns. 27-28]{breiter+2012} derived for the $\,\mathbb{B}\,$ matrix a more general expression that was appropriate for a triaxial body. Our formulae (\ref{37}) agree with that expression in the oblate case where $\,h_1 = 1~$ and $~h_2 = h\,$.}
 \begin{subequations}
 \ba
 B_{ \rm  11}=\,\left(\Omega_3^2+\Omega_2^2\right)\,-\gamma_1~~,~~B_{ \rm  22}\,=\,\left(\Omega_3^2+\Omega_1^2\right)\,-\gamma_2~~,~~
 \label{37a}
 \ea
 \ba
 ~~B_{ \rm  33}\;=\;\Omega_1^2\;+\;\Omega_2^2\;-\;\gamma_3~~,~~~~B_{ \rm  12}=\,-\,\Omega_{ \rm  1}\,\Omega_{ \rm  2}~~,~~
 \label{37b}
 \ea
 \ba
 B_{ \rm  13}\;=\;-\;\frac{2\,\Omega_{ \rm  1}\,\Omega_{ \rm  3}}{1\,+\,h_2}~~~,~~~~B_{ \rm  23}=\,-\,\frac{2\,\Omega_{ \rm  2}\,\Omega_{ \rm  3}}{1\,+\,h^{ \rm  2}}~~~,~~
 \label{37c}
 \ea
 \ba
 B_{ \rm  21}\,=\,B_{ \rm  12}~\,,~~B_{ \rm  31}=\,h^2\,B_{ \rm  13}~\,,~~B_{ \rm  32}=\,h^2\,B_{ \rm  23}~\,.~\,
 \label{37d}
 \ea
 \label{37}
 \end{subequations}

 \subsection{Generic difficulties} \label{generic}

 There are two major difficulties associated with the afore-derived expression (\ref{36}). These difficulties are generic, in that they are encountered, no matter what geometric shape is chosen --- ellipsoid, prism, octahedron, whatever. Here we describe these difficulties, and explain in brief how they should be resolved. A detailed treatment of these issues is presented in Appendix \ref{difficulties}.

 \subsubsection{The oscillating component $\tilde{\bf{b}}_{ \rm  gr}$ of self-gravitation}

 Was it really acceptable to neglect in the equation (\ref{36}) the oscillating component $\tilde{\bf{b}}_{ \rm  gr}$ of self-gravitation? It is tempting to give to this question an affirmative answer, vaguely basing it on the fact that deformations are ``small". \footnote{Recall that we assumed \textit{ab initio} that the deformations are small enough to justify the transition from the true equations of motion, (\ref{6}), to their simplified version (\ref{8}).}

 Unfortunately, the argument of smallness fails to work. To understand this, consider a similar situation emerging in the theory of tides, where deformation-caused changes in the gravitational potential cannot be omitted. Indeed, by calculating those small changes and comparing them with the tide-raising potential, we find the holy grail of the tidal theory, the dynamical Love numbers $\,k_l\,$ and phase lags $\,\epsilon_l\,$ with $\,l\geq2\,$. Thereafter, the tidal dissipation rate is calculated as a series where each term is proportional to $\,k_l\,\sin\epsilon_l\,$, see \cite{pealecassen1978} and \cite{efroimskymakarov2014}. Evidently, the neglect of the small changes of shape implies setting $\,k_l=0\,$ which, in its turn, renders a zero dissipation rate, a nonsensical outcome.

 However, there exists an alternative, though equivalent, way of calculating the dissipated power. It implies integration over the body (and averaging over time) of a product of the stress by the time derivative of the strain. This method is explained briefly in Section \ref{calculation} below, and in more detail in Appendix \ref{difficulties}. In Section \ref{AppendixD3} of the Appendix, it is demonstrated that the so-arranged calculation renders the damping rate as a sum over the harmonics $\,\chi\,$ involved, with each term in this sum proportional to $\,|\,\bar{J}(\chi)\,|\,\sin\delta(\chi)\,$. Here $\,\bar{J}(\chi)\,$ is the 
 complex compliance of the medium at the frequency $\,\chi\,$, while $\,\delta(\chi)\,$ is the phase lag of the strain relative to the stress at this frequency. This lag is defined through
 \begin{equation}
 \bar{J}(\chi)~=~|\,\bar{J}(\chi)\,|~\exp[\,-\,i\,\delta(\chi)\,]\;.
  \end{equation}
 The neglect of the small variations of shape does $\,${\it{not}}$\,$ nullify $\,|\,\bar{J}(\chi)\,|\,$, because this quantity is local and is defined by the rheology of the material (not by the shape of the body). Accordingly, within this formulation, it is possible to ignore the small changes of the shape and the ensuing oscillating component $\,\tilde{\bf{b}}_{ \rm  gr}\,$ of self-gravitation.

 \subsubsection{The constant part of the stress} \label{constantstress}

 It is not {\it{a priori}} evident whether in our further development we should remove from the expressions (\ref{36} - \ref{37}) the time-independent part (called prestressing). This part comprises the mean centrifugal force and the constant component $\,\overline{\bf{b}}_{ \rm  gr}\,$ of the gravity force. To calculate the dissipated power, we multiply the overall body force $\bold{f}$ by the oscillating deformation velocity $\,{\bf{\dot{u}}}\,$, integrate the result over the volume, and time-average it. Na{\"{i}}vely, the constant part of the body force drops out after time-averaging; and only the harmonics of the force, consonant with the harmonics of the velocity, are left. In reality, however, averaging does \textit{not} eliminate prestressing, because the said calculation should be performed in the Euler coordinates. In those coordinates, prestressing is $\,${\it{not}}$\,$ constant in time. (It is constant only when expressed via the Lagrange coordinates!) Whether a correction rendered by this circumstance is small or not is a highly nontrivial issue.

 An alternative way of calculating the dissipated power is to integrate over the body a product of the stress by the time derivative of the ensuing strain. Once again, the constant part of the stress should drop out after time-averaging; and only the harmonics of stress multiplied by the appropriate harmonics of strain rate should remain. In reality, accurate calculation of the stress implies solution of the full boundary-value problem, with the so-called compatibility conditions brought in. This solution includes all information about the perturbation, including prestressing.

 In Appendix \ref{AppendixD3} we present arguments justifying the neglect of prestressing when the effective viscosity $\,\eta\,$ is not too low:
 \begin{equation}
 \eta\;\gg\;\frac{\textstyle{8\,\pi\,G\,\rho^2\,R^2}}{\textstyle{57\;\chi}}\;,
 \label{criterion}
 \end{equation}
 where $\,\chi\,$ is the lowest forcing frequency in the spectrum. For a dynamically oblate rotator, it is the wobble frequency
 \be
 \chi\;=\;\omega\;,
 \ee
 For asteroids of viscosity higher than the threshold (\ref{criterion}), prestressing may be ignored.

 The wobble frequency $\,\omega\,$ is of the order of 1/hr to 1/day, i.e., $\,10^{ \rm  -5}\,-\,10^{ \rm  -3}\,$ Hz. The density is  $\,\rho\,\approx\,2\times 10^{ \rm  3}\,$ kg/m$^3\,$.
 So prestressing may be omitted for
 \begin{equation}
 \eta\;\gg\;(10^{ \rm  -1}\;\mbox{---}\,\;10^1)~R^2\;,
 \end{equation}
 where $\,R\,$ is given in meters. For example, in the case of a twenty-kilometer-sized asteroid ($\,R=10^4\,$ m), the effective viscosity must exceed $\,10^7 \;\mbox{---}\,\; 10^9\,$ Pa$\,\cdot\,$s$\,$, in order for prestressing to be unimportant.

 \subsection{Inelastic relaxation of a precessing rotator}

 Three time scales are present in our problem. The fastest is the period of the instantaneous spin. The intermediate one is associated with precession. The slowest time scale is that of precession relaxation.

 Precession yields deformation at several frequencies. As we mentioned in Section \ref{3.2.3}, for an oblate rotator they are two: $\,\chi_1\,=\,\omega\,$ and $\,\chi_2\,=\,2\,\omega\,$; while for a triaxial body we get an infinite number of forcing frequencies that are harmonics of the lowest frequency: $\,\chi_n\,=\,n\,\chi_1\,$. In the latter situation, most dissipation is taking place at several lowest overtones, so we can assume that the important modes do not differ from $\,\chi_1\,$ by more than an order of magnitude, or so.

 Over a period of a forcing frequency, the internal energy $\,E\,$ first borrows a portion from $\,T_{ \rm  kin}\,$ (the compression part of the cycle),
 and then returns most of that portion back to $\,T_{ \rm  kin}\,$ (the expansion part of the cycle) -- see the equation (\ref{dissipation}) in Appendix \ref{AppendixA}. A small difference between the loan and the return (given by the equation \ref{ged} in Appendix \ref{AppendixA}) gets dissipated. When the body is in a thermodynamical equilibrium with the environment, this small difference is completely radiated away, so the internal energy $\,E\,$ regains its original value by the end of the cycle. Thence the internal energy $\,E\,$ averaged over the cycle stays unchanged also:
 \begin{equation}
 \langle E \rangle~=~\mbox{const}\;,
 \label{E}
 \end{equation}
 $\langle\,.\,.\,.\,\rangle\,$ denoting average over the intermediate (precession-related) time scale.

 \subsubsection{Energy balance}

 Over a time interval $\,\Delta t\,$, the energy balance of a body is given by
 \begin{equation}
 \Delta W\,+~\Delta {\cal{Q}}~=~\Delta E~+~\Delta T_{ \rm  kin}\;,
 \end{equation}
 where $\,W\,$ is the external mechanical work, $\,{\cal{Q}}\,$ is the heat transferred to the body, $\,E\,$ is the internal energy, while $\,T_{ \rm  kin}\,$ is the kinetic energy of the body as a whole. Neglecting the exterior mechanical influences, we assume:
  \begin{equation}
 \Delta W~=~0\;.
 \label{W}
  \end{equation}
 According to the equation (\ref{E}),
  \begin{equation}
 \langle \Delta E \rangle ~=~0\;,
 \label{E1}
  \end{equation}
 with averaging over the forcing period (the precession timescale). So we are left with
  \begin{equation}
 \langle \Delta T_{ \rm  kin}\rangle\,=~\langle\Delta {\cal{Q}}\rangle\;\;.
  \end{equation}
 Assume that the body radiates away all the heat it obtains from insolation.$\,$\footnote{~Here we neglect the YORP effect. However, there exist situations where rotational dynamics is defined by combination of YORP with inelastic relaxation -- see \cite{breiter+2011}, \cite{breitermurawiecka2015}, and references therein.} So the overall heat influx, $\,\langle\Delta {\cal{Q}}\rangle\,$, is negative, for it is equal to minus the heat dissipated by friction:
 \begin{equation}
 \langle\Delta {\cal{Q}}\rangle~=~-~\langle P\rangle~<~0\;.
 %\label{}
  \end{equation}
 The former and the latter formulae, together, yield:
  \begin{equation}
 \langle\dot{T}_{ \rm  kin}\rangle\,=~-~\langle P\rangle~=~-~\sum_n\langle{P}(\chi_n)\rangle\;,
 \label{43}
  \end{equation}
 where the total power $\,P\,$ comprises the damping rates $\,P(\chi_n)\,$ at all the frequencies $\,\chi_n\,$ of the deformation spectrum.

 The rate of precession relaxation can be linked to the rate $\,\dot{T}_{ \rm  kin}\,$ of kinetic energy loss~---~which coincides with the total dissipated power $\,P\,$. The dissipation rate at each involved frequency, $\,P(\chi_n)\,$, can be linked to the phase lag between the appropriate harmonics of stress and strain. These lags are determined by the material rheology, spin state, and self-gravitation of the rotator. This way, the speed of precession relaxation is defined by rheology, spin and self-gravitation.

  \subsubsection{Measure of precession}

 Deviation of an unsupported rotator from the minimal-energy state can be measured with $\theta$, the angle between the major-inertia axis
 and the angular momentum vector ${\bf{J}}\,$:
   \begin{equation}
 J\;\cos\theta\;=\;I_3\,\Omega_3\;.
 \label{A1}
  \end{equation}
  For an oblate rotator, the angle $\,\theta\,$ stays constant over a precession cycle ({\it{adiabatically}}$\,$ constant, as was explained in the preceding subsection). So $\,d\theta/dt\,$ can be employed as the rate of precession relaxation of an oblate body. This rate can be written down as
 \begin{equation}
 \frac{d \theta}{dt}~=~\frac{d \theta}{dT_{ \rm  kin}}~\frac{dT_{ \rm  kin}}{dt}\;.
 \label{45}
 \end{equation}
 As was demonstrated in \cite{efroimskylazarian2000},
 \begin{equation}
 \frac{d \theta}{dT_{ \rm  kin}}~=~\bigg[\frac{J^2}{I_3} \bigg(\frac{I_3}{I_1} - 1\bigg) \sin \theta \cos \theta \bigg]^{ \rm  -1}\;.
 \end{equation}
 Bringing in the expression (\ref{43}) for the energy leakage, we write the relaxation rate as
 \begin{equation}
 \frac{d \theta}{dt}~=\;-\;\bigg[\frac{J^2}{I_3} \bigg(\frac{I_3}{I_1} - 1\bigg) \sin \theta \cos \theta \bigg]^{ \rm  -1}\sum_n\langle\,P(\chi_n)\,\rangle\;.
 \label{47}
 \end{equation}
 While time averaging is implied on both sides, we put the angular brackets only on the right-hand side, not on the left. This is all right, since the averaging is performed over one or several precession periods, while appreciable changes in $\,\theta\,$ accumulate over much longer timescales. Variations of $\,\theta\,$ over shorter times are minuscule. The damping time is then
 \begin{equation}
 T\,=\;-\;\int_{ \rm  \theta_{ \rm  0}}^{ \rm  \theta_{ \rm  f}} \bigg[\,\frac{J^2}{I_3}~\bigg(\frac{I_3}{I_1}~-~1\bigg)\;\sin \theta\;\cos \theta\,\bigg]\;\frac{1}{\langle P\rangle}\;d\theta\;,
 \label{eqdamp}
 \end{equation}
 where the quadrature covers the evolution of the nutation angle from its initial value $\,\theta_{ \rm  0}\,$ to the final value $\,\theta_{ \rm  f}\,$.

 \subsubsection{Calculation of the dissipated power, by switching from vectors to tensors} \label{calculation}

 Whether we have to employ the expression (\ref{47}) or (\ref{eqdamp}), our main challenge is to compute the dissipation rate $\,\langle P\rangle\,$.
 Fundamentally, the dissipated power is given by \footnote{~In the presence of YORP, there emerges a traction force $\,{\bf{t}}\,$ acting on a part of the surface. In this situation, the second term in the formula ({\ref{51}}) should be kept.
 % The YORP forces not only create a torque but also make an extra contribution into dissipation, by interacting with the surface deformation.
 }
 \begin{equation}
 P(t)~=~\int_V \rho~\textbf{b}(\textbf{x},t) \cdot {\bf{\dot{u}}}(\textbf{x},t)~dV
 + \int_S \textbf{t}(\textbf{x},t)
 % ^{ \rm  (\textbf{n})}
 \cdot {\bf{\dot{u}}}(\textbf{x},t)~dS
 \label{51}
 \end{equation}
 $\textbf{b}$ being the body force per unit mass, $\textbf{t}$ denoting the traction force, $\rho$ being the density (assumed constant), $\textbf{u}$ standing for the displacement, and overdot signifying time derivative. The coordinates $\textbf{x}$ are those of Euler, so $dV \equiv d^3 \textbf{x}$ is a Eulerian (deformed) element of volume, while $\,dS\,$ is an element of the deformed surface.
  As no traction is applied on the surface, the traction-related part vanishes.

  The definition of strain, $~e_{ \rm  ij}\,=\,\frac{\textstyle 1}{\textstyle 2}\,\left(\frac{\textstyle\partial u_i}{\textstyle\partial x_j}\,+\,\frac{\textstyle\partial u_j}{\textstyle\partial x_i}\right)~$, and the relation $\rho\,\bf{b} = -\,\nabla \mathbf{\sigma}$ linking the exterior force to the stress will help us to write the volume integral as
  \begin{eqnarray}
 P(t) & = & \int_V \rho~\textbf{b}(\textbf{x},t) \cdot {\bf{\dot{u}}}(\textbf{x},t)~dV \; \nonumber\\
      & = &   -\;\int_V \frac{\rho}{2}\,\left[\,\left(\,\partial_i\,\sigma_{ \rm  ij}\,\right)\, {\dot{u}}_j\,+\,\left(\,\partial_j\,\sigma_{ \rm  ji}\,\right)\,{\dot{u}}_i \,\right]~dV\;,\quad\qquad
 \label{52a}
 \end{eqnarray}
 A subsequent integration by parts yields a surface term (vanishing due to the boundary condition $\,\sigma{\bf{\hat{n}}}=0\,$) and a volume integral that can be
 simplified by using $\,\sigma_{ \rm  ji}=\sigma_{ \rm  ij}\;$:
  \begin{eqnarray}
 P(t)~&=&~\int_V \frac{\rho}{2}\,\left[\,\sigma_{ \rm  ij}\,\left(\,\partial_i\, {\dot{u}}_j\,\right)\,+\, \sigma_{ \rm  ji}\,\left(\,\partial_j\,{\dot{u}}_i\,\right)\,\right]~dV \nonumber \\
 ~&=&~\int_V \sigma_{ \rm  ij}(\textbf{x},t)~\dot{e}_{ \rm  ij}(\textbf{x},t)~dV~~. \quad\qquad
 \label{52b}
 \end{eqnarray}
 Averaging over the intermediate (precession-related) timescale, we arrive at \footnote{~Rigorously speaking, taking the volume element $\,dV\,$ out of the averaging is legitimate only when the integration is performed over the Lagrange variables. In our case, however, we are dealing with the Euler (time-dependent) coordinates and a deformable volume element. The deformable volume element $\,dV\,$ is related to its Lagrange (undeformable) counterpart $\,dV_{ \rm  L}\,$ by the relation $\,dV\,=\,J\,dV_L\,$. As the Jacobian is given by $\,J\,=\,1\,+\,\nabla{\bf{u}}\,$, where $\,{\bf{u}}\,$ is the displacement field, we see that the neglect of the Jacobian entails a higher-order error. In a leading-order calculation, the Jacobian may be set unity, wherefore our taking of $\,dV\,$ outside averaging in equation (\ref{53b}) is acceptable. For a more detailed explanation, see the third paragraph of Section 5.1 in  \citet{efroimskymakarov2014}.}
 %  \bs
 \begin{eqnarray}
 P&=&\left\langle\;\int_V \sigma_{ \rm  ij}(\textbf{x},t)~\dot{e}_{ \rm  ij}(\textbf{x},t)~dV\;\right\rangle \nonumber \\
         ~&=&~
 \int_V \left\langle\;   \sigma_{ \rm  ij}(\textbf{x},t)~\dot{e}_{ \rm  ij}(\textbf{x},t)  \;\right\rangle~dV
 ~_{ \rm  \textstyle{_{ \rm  \textstyle{_{ \rm  \textstyle{.}}}}}}
 \label{53b}
 \end{eqnarray}
 On the left-hand side, we removed the time dependence but did not bother to write the angular brackets. Here and hereafter, we shall omit these brackets when dealing with the power or kinetic energy, implying that $\,P\,$ and $\,T_{ \rm  kin}\,$ denote the $\,${\it{averaged}}$\,$ quantities.

 \section{Dissipation rate and nutation relaxation of viscoelastic oblate spheroids and rectangular prisms}

 We are now prepared to compute the dissipation rate and the nutational damping time for bodies of two different shapes. The bodies are assumed to be homogeneous and to consist of a Maxwell material. This means that the volumetric strain rate $\,\stackrel{\centerdot}{\boldsymbol{e}}^{ \rm  (V)}\,$ is related to the volumetric stress $\,{\boldsymbol{\sigma}}^{ \rm  (V)}\,$ as
 \bs
 \ba
 \stackrel{\centerdot}{\boldsymbol{e}}^{ \rm  (V)}=\,\frac{1}{3\,K}\;\stackrel{\centerdot}{\boldsymbol{\sigma}}^{ \rm  (V)}+\,\frac{1}{2\,\zeta}\;{\boldsymbol{\sigma}}^{ \rm  (V)}\,,
 \ea
 $K\,$ and $\,\zeta\,$ being the bulk rigidity and bulk viscosity. The deviatoric
 % strain rate $\,\stackrel{\centerdot}{\boldsymbol{e}}^{ \rm  (D)}\,$ is related to the deviatoric stress $\,{\boldsymbol{\sigma}}^{ \rm  (D)}\,$
 counterparts of these tensors are linked
 via
 \ba
 \stackrel{\centerdot}{\boldsymbol{e}}^{ \rm  (D)}=\,\frac{1}{2\,\mu}\;\stackrel{\centerdot}{\boldsymbol{\sigma}}^{ \rm  (D)}+\,\frac{1}{2\,\eta}\;{\boldsymbol{\sigma}}^{ \rm  (D)}
 \,,
 \ea
 \es
 $\mu\,$ and $\,\eta\,$ being the shear rigidity and the shear viscosity, correspondingly. For details, see Appendix \ref{AppendixC} below.

\subsection{Oblate ellipsoid\label{4.1}}

 \cite{sharma+2005} calculated the stress tensor in an {{elastic}} oblate spheroid with a Poisson's ratio $\,\nu=1/4\,$. Using those expressions, we shall keep only the time-dependant part thereof (see the discussion in Section \ref{constantstress}). Since the stress does not depend on the rheological parameters, we can use that result also for a viscoelastic spheroid: $\,\sigma_{ \rm  ij}^{ \rm  (viscoelastic)}\,=\,\sigma_{ \rm  ij}^{ \rm  (elastic)}\,$.

 We separate the stress into its volumetric and deviatoric parts:
 \be
 \nonumber
 \boldsymbol\sigma = \boldsymbol\sigma^{ \rm  (V)} + \boldsymbol\sigma^{ \rm  (D)}\;,
 \ee
 where the volumetric stress is defined by
 \begin{equation}
 \sigma^{ \rm  (V)}_{ \rm  ij}\,=~\frac{\delta_{ \rm  ij}}{3}\,\left(\sigma_{ \rm  11} + \sigma_{ \rm  22} + \sigma_{ \rm  33}\right) ~.
 \end{equation}
 Then we find, separately, the volumetric and deviatoric parts of the strain. The volumetric strain is easily obtained through
 \be
 \boldsymbol e^{ \rm  (V,viscoelastic)}\,\approx\;\boldsymbol e^{ \rm  (V,elastic)}\,=\;\frac{1}{3 K} \boldsymbol\sigma^{ \rm  (V)}~.
 \ee
 The reason for such a simplification is the following. While the bulk (volumetric) rigidity modulus $\,K\,$ is only slightly larger than its shear (deviatoric) counterpart $\,\mu\,$,  \footnote{~For the value $\,\nu=1/4\,$, we get: $\;{\textstyle K}\,=\,\frac{\textstyle{2}}{\textstyle{3}}\,\frac{\textstyle{1+\nu}}{\textstyle{1-\nu}}\,{\textstyle \mu}\;=\;\frac{\textstyle{10}}{\textstyle{9}}\,{\textstyle \mu}\,$.} it is known in continuum mechanics that the bulk (volumetric) viscosities $\,\zeta\,$ for most realistic viscoelastic substances are hundreds to thousands of times larger than the shear (deviatoric) viscosities $\,\eta\,$. Hence, for most viscoelastic media, the viscous input into the volumetric strain is much smaller than the elastic input.$\,$\footnote{~This is not so for planetary mantles, because the presence of partial melt changes the situation drastically. As melt fraction increases from zero to the critical melt fraction, rapid decreases occur in both the shear and bulk viscosities. While the shear viscosity falls by a factor of five,
 the bulk viscosity can fall by a factor from hundreds to thousands, and becomes only about twice larger than the shear viscosity when the melt fraction becomes
 high~---~see \citet[Fig 9]{takei}.}

 To find the deviatoric part of the strain,  $\,\boldsymbol e^{ \rm  (D)}\,$, we assume that the rheology is Maxwell, and resort to a wonderful theorem known as the correspondence principle. This calculation is presented in Appendix \ref{AppendixE}. We then obtain the complete strain tensor as
 \be
 {\boldsymbol{e}}\;=\;\boldsymbol{e}^{ \rm  (V)}\,+\;\boldsymbol{e}^{ \rm  (D)}\,.
 \ee
 The expressions for the components of this tensor are listed in Appendix \ref{shapes}.

 The mean dissipated power is found as an (averaged over time) integral over the volume of the strain rate times the stress, as described in the equation (\ref{53b}). The volume integration is performed in spherical coordinates according to the formula (\ref{27}). The algebra is straightforward.

 With aid of the shape parameter $\,h\,$ and the nominal angular rate $\,\tilde{\omega}_3\,$ defined as
 \ba
 h\;\equiv\;\frac{c}{a}\;\,,\qquad\tilde{\omega}_3\,\equiv\;\frac{J}{I_3}\;,
 \ea
 the expression for the generated power assumes the form of
 \begin{equation}
 \begin{split}
 P = & \frac{a^7\;\tilde{\omega}_3^4\;\rho^2}{\eta}  \frac{\pi}{(h^2\,+\,1)^4} \\
\times &  \bigg[ \frac{16}{315} h \frac{384 h^8 + 960 h^6 + 1900 h^4 + 1650 h^2 + 1125}{(8 h^4 + 10 h^2 + 15)^2} \sin^4 \theta \\
& + \frac{32}{315} (h^2 + 1)^2 h^5  \frac{(1050 h ^4  + 2015 h^2 + 507) }{(20 h^2 + 13)^2} \sin^2 \theta  \cos^2 \theta  \bigg]
 \label{eq1}
\end{split}
\end{equation}

  As we see, only terms containing the viscosity $\,\eta\,$ survive the time average. No terms with the bulk or shear rigidity show up in the answer. This parallels a similar situation with tidal interaction of asteroidal binaries \citep{efroimsky2015}. As explained in \citet{efroimsky2015}, this outcome should not be misinterpreted as total irrelevance of rigidity.
 (After all, we need some rigidity to sustain the oblate shape. So we cannot extend our treatment to liquid bodies with no rigidity.)

 Plugging the formulae (\ref{23}) into the formula (\ref{47}), we obtain the relaxation rate:
 \begin{equation}
 \frac{d\theta}{dt}\;=\;-\;\frac{d\theta}{dT_{ \rm  kin}}\;P\;=\;-\;\frac{2\;\rho\;a^5}{5\;J^2\;\sin \theta\;\cos \theta}\;\frac{\textstyle 1\,+\,h^2}{\textstyle 1\,-\,h^2}\;P\;.
%= \frac{3 (1+h^2)}{16 \rho a^5 h (1-h^2) \tilde{\omega}_3^2 \cos(\theta) \sin(\theta)  } P
 \label{q}
 \end{equation}
 Inserting therein the expression (\ref{eq1}) for the power, and integrating, we obtain the nutational damping time:
\begin{equation}
\begin{split}
T & = \frac{\eta }{a^2 \rho  \, \tilde{\omega}_3^2}  \int_{ \rm  \theta_0}^{ \rm  \theta_f}   (h^2 + 1)^3 (1-h^2)   \cos \theta            \\
\times & \bigg[ \frac{4}{21} h^4 (h^2 + 1)^2 \frac{1050 h^4 + 2015 h^2 + 507}{(20 h^2 + 13)^2} \sin \theta \cos^2 \theta \\
 & \frac{2}{21} \frac{384 h^8 + 960 h^6 + 1900 h^4 + 1650 h^2 + 1125}{(8 h^4 + 10 h^2 + 15)^2} \sin^3 \theta \bigg]^{ \rm  -1} d\theta 
\label{eq2}
\end{split}
\end{equation}

 We have written down this expression in a form that would facilitate comparison with the result from \citet{breiter+2012}. For convenience, below we cite their result in the equation (\ref{eq5}). Our $\,T\,$ should be compared with the quantity $\,T_3\,$ in \citet{breiter+2012}.

 The dependence of the relaxation time $\,T\,$ upon the shape parameter $\,h=c/a\,$, as given by our expression (\ref{eq2}), is presented graphically in the upper panel of Figure \ref{fig1}. There, three curves depict the evolution of the nutation angle $\,\theta\,$ from three initial values $\,\theta_0\,=\,20^\circ\,$,$\,45^\circ\,$,$\,85^\circ\,$ down to the same final value of $\,\theta_f\,=\,5^\circ\,$.

 The dependence of the normalised dimensionless damping time $T_{ \rm  dim.less} = T \,\frac{\textstyle \rho a^2 \tilde{\omega}_3^2}{\textstyle \eta}$ on $h$ in the range [0.5,1] can be satisfactorily fitted with a polynomial, such that the damping time in that range reads
 \begin{equation}
 \begin{split}
 T & =  \frac{\textstyle \eta}{\textstyle \rho \, a^2 \, \tilde{\omega}_3^2} \, T_{ \rm  dim.less} \\
   &  = \frac{\textstyle \eta}{\textstyle \rho \, a^2 \, \tilde{\omega}_3^2} \, (d_1 h^4 + d_2 h^3 + d_3 h^2 + d_4 h + d_5) \, .
 \label{eq7}
\end{split}
 \end{equation}
 where the polynomial coefficients are listed in Table \ref{tab}. For example, for an initial $\theta$ = 20\textdegree, we have $T_{ \rm  dim.less}= 1.15$ for $h$ = 0.9, and $T_{ \rm  dim.less}= 3.13$ for $h$ = 0.8.

\begin{table}
 \begin{center}
\begin{tabular}{l|rrr}
    & \multicolumn{3}{c}{$\theta$ initial}\\
      &20\textdegree  & 45\textdegree & 85\textdegree\\
\hline
$d_1$ & 479.854 &  388.48 & 405.69\\
$d_2$ & -1724.3 & -1428.77 & -1477.09 \\
$d_3$ & 2363.95 & 2026.11  &  2068.67 \\
$d_4$ & -1479.81 &-1331.96  & -1346.29 \\
$d_5$ & 360.37 &  346.184 & 349.063
 \end{tabular}
 \end{center}
 \caption{Coefficients of the fitting polynomial in the formula for the damping time of a viscoelastic oblate ellipsoid (\ref{eq7}) in the range $h$ = [0.5,1].}
 \label{tab}
\end{table}

\subsection{Dynamically oblate rectangular prism}

We use the elastic solution of the stress tensor for an rectangular prism $2a \times 2a \times 2c$, with $a>c$, computed by \cite{efroimskylazarian2000}, \cite{efroimsky2001}. The stress tensor is listed in Appendix \ref{AppendixE}. Application of the equivalence principle renders the viscoelastic strain tensor described in Appendix \ref{shapes}. For a rectangular prism, the moments of inertia are
 \begin{equation}
 I_3 = \frac{16}{3}\;\rho\;a^5\;h \quad, \quad  I_1\,=\;I_2\,=\;I\;=\;\frac{I_3}{H}\;,
 \end{equation}
 with $\,H\,=\,\frac{\textstyle 2}{\textstyle 1+h^2}\,$ and $\,h=c/a\,$. The power reads as
 \ba
 \nonumber
 P = \frac{a^7~\rho^2~\tilde{\omega}_3^4}{\eta}~\frac{h}{(1\,+\,h^2)^4}
 \;\bigg( \frac{1328}{135} \sin^4 \theta\qquad\qquad\quad\quad\\
 \nonumber\\
  + \frac{2816}{45} h^4 \sin^2 \theta  \cos^2 \theta  \bigg)\;.\qquad\quad
 \label{eq3}
 \ea
The rate of change of the nutation angle is
\begin{equation}
\frac{d\theta}{dt}\;=\;-\;\frac{d\theta}{dT_{ \rm  kin}}\;P\;=\;-\;\frac{16\;\rho\;a^5}{3\;J^2\;\sin \theta\;\cos \theta}\;\frac{h\;(1\,+\,h^2)}{1\,-\,h^2}\;P\;.
%= \frac{3 (1+h^2)}{16 \rho a^5 h (1-h^2) \tilde{\omega}_3^2 \cos(\theta) \sin(\theta)  } P
\label{g}
\end{equation}
Reversing this equation and integrating over $\theta$, we obtain the nutation damping time
 \begin{equation}
 T\,=\;-\;\frac{\eta }{a^2\,\rho\,\tilde{\omega}_3^2}\,\int_{ \rm  \theta_0}^{ \rm  \theta_f}
 \frac{(1+h^2)^3 (1-h^2) \cos \theta}{ \frac{\textstyle 83}{\textstyle 45} \sin^3 \theta + \frac{\textstyle 176}{\textstyle 15} h^4 \sin \theta  \cos^2 \theta } d\theta \;.
 \label{eq4}
 \end{equation}
 This dependence of the relaxation time $\,T\,$ on the shape parameter $\,h=c/a\,$ is plotted in the lower panel of Figure \ref{fig1}. There, three curves present the evolution of the nutation angle $\,\theta\,$ starting from three different initial values $\,\theta_0\,=\,20^\circ\,$,$\,45^\circ\,$,$\,85^\circ\,$ and ending at the final value of $\,\theta_f\,=\,5^\circ\,$.

 \section{Results and discussions}

 \begin{figure}
  \text{\hspace{1.0cm}}   \resizebox{9cm}{!}{\includegraphics [angle=0,width=\textwidth] {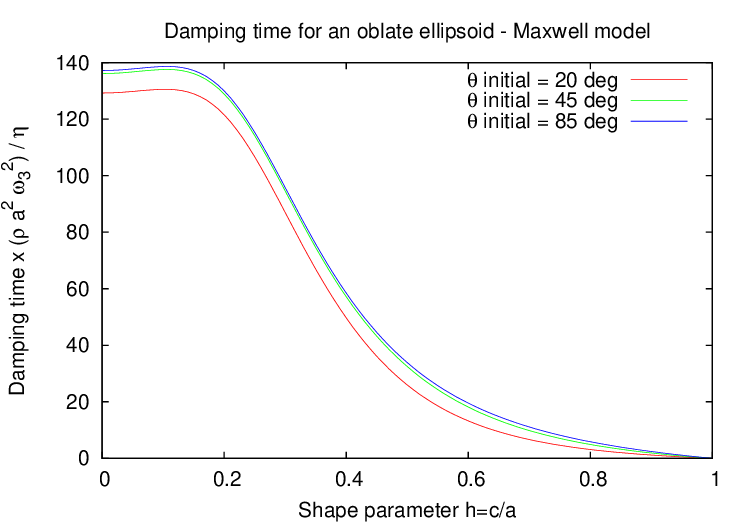}}
  \text{\hspace{1.0cm}}   \resizebox{9cm}{!}{\includegraphics [angle=0,width=\textwidth] {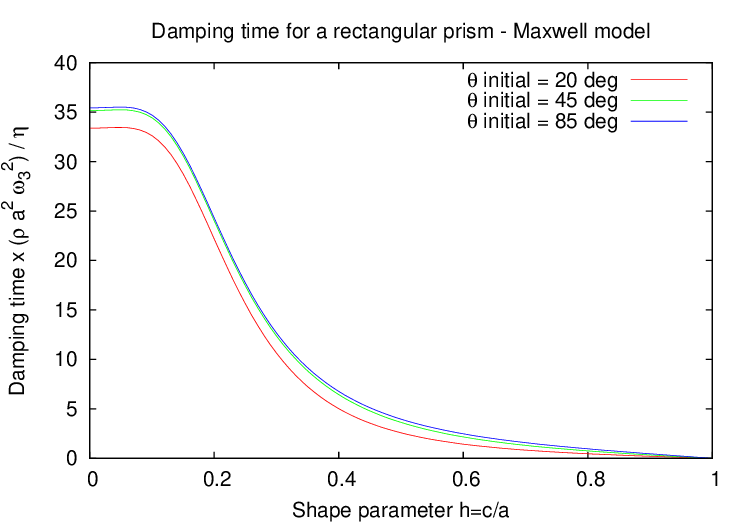}}
~\\~\\
 \caption{Normalised damping time for the Maxwell model, for an oblate ellipsoid (eqn. \ref{eq2}) and a rectangular prism (eqn. \ref{eq4}), presented as a function of the shape parameter $\,h=c/a\,$. The three curves depict the evolution of the nutation angle $\,\theta\,$ from three different initial values $\,\theta_0\,$ (indicated in the figures) to the same final value $\,\theta_f\,=\,5^\circ\,$.}
\label{fig1}
\end{figure}

\begin{figure}
 \text{\hspace{1.0cm}}   \resizebox{9cm}{!}{\includegraphics [angle=0,width=\textwidth] {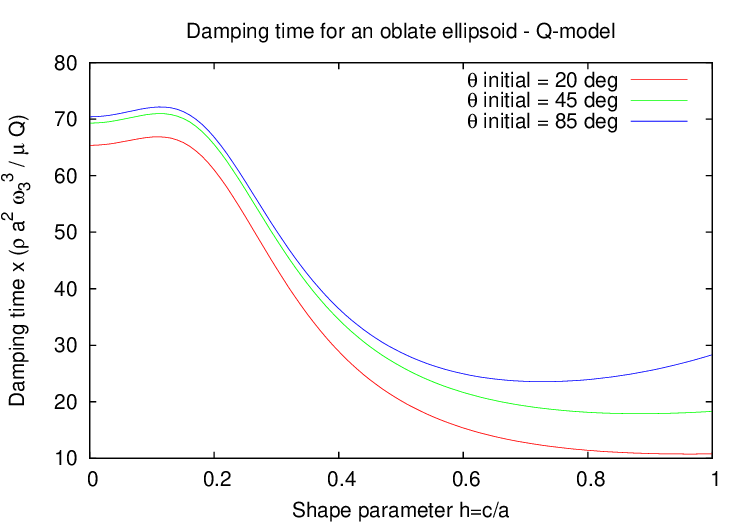}}
 \text{\hspace{1.0cm}}   \resizebox{9cm}{!}{\includegraphics [angle=0,width=\textwidth] {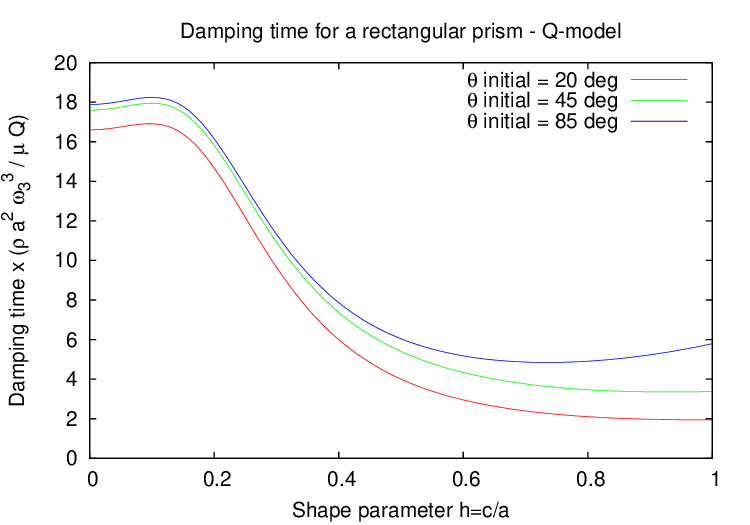}}
~\\~\\
 \caption{Normalised damping time for the empirical model based on an $\,${\it{ad hoc}}$\,$ quality factor $\,Q\,$, for an oblate ellipsoid (eqn. \ref{eq5}) and a rectangular prism (eqn. \ref{eq6}), presented as a function of the shape parameter $\,h=c/a\,$.}
\label{fig2}
\end{figure}

 It remains to compare the damping time (\ref{eq2}) of an oblate spheroid with the damping time (\ref{eq4}) of a dynamically oblate rectangular prism. The functional forms of the damping times' dependencies on the parameter $\,h=c/a\,$ are not fundamentally different between the two shapes. Accordingly, similar are the resulting curves depicting $\,T\,$ as a function of $\,h\,$.

 Figure \ref{fig1} shows the normalised dimensionless damping time $\,T\,a^2\,\rho\,\tilde{\omega}_3^2\,\eta^{ \rm  -1}\,$ as a function of $\,h\,$. The upper panel presents the dependence computed from the equation (\ref{eq2}) for an oblate spheroid. The lower panel exhibits the dependence obtained from the equation (\ref{eq4}) for a prism.
 In both cases, we evaluated the integral over $\,\theta\,$ numerically, with the final angle $\,\theta_f\,=\,5^\circ\,$ and with the initial angle assuming three different values: $\,\theta_0\,=$ ($20^\circ,\,45^\circ,\,85^\circ$). In agreement with an observation made in \cite{efroimsky2001}, it is possible to deduce from the above equations that in the vicinity of the principal spin state the relaxation rate becomes exponentially
 slow.$\,$\footnote{~To demonstrate this, plug the formula (\ref{eq1}) into the equation (\ref{q}), for a spheroid; or plug the formula (\ref{eq3}) into the equation (\ref{g}), for a prism. On both occasions, expansion of the resulting expression for $\,d\theta/dt\,$ into a Taylor series near $\,\theta=0\,$ leads to $\;{d\theta}/{dt} \propto\,-\;\theta\,$.} Hence the necessity to choose a non-zero value $\,\theta_f\,$ for the final nutation angle.

 The shapes of these functions are similar to the $\,h-$dependence of the ``shape factor", presented in \cite{sharma+2005} and \cite{breiter+2012}.

 We see from the plots that relaxation is faster when the precession angle $\,\theta\,$ is large. Formally, $\,d\theta/dt\,$ is infinite at $\,\theta = \pi/2\,$ (see Figure \ref{fig4}). Thus, during the evolution, most of the relaxation time is spent at moderate to low values of $\,\theta\,$, explaining the small difference between the curves with initial $\,\theta_0 = $ $45^\circ\,$ and $\,85^\circ\,$.

\begin{figure}
%\begin{center}
 \text{\hspace{1.0cm}}   \resizebox{9cm}{!}{\includegraphics [angle=0,width=\textwidth] {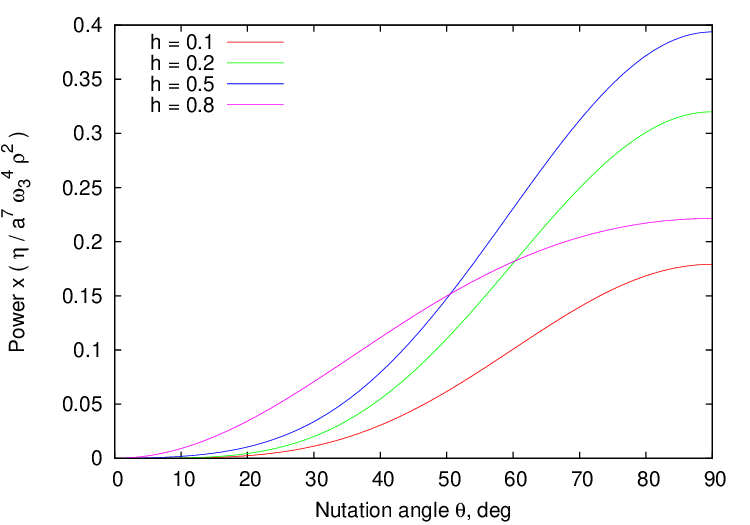}}
~\\
~\\
\caption{Normalised power $\;P/(a^7\,\omega_3^4\,\rho^2/\eta)\;$ for a viscoelastic oblate ellipsoid (eqn. \ref{eq1}), for various values of the shape parameter $\,h=c/a\,$.}
\label{fig3}
%\end{center}
\end{figure}

\begin{figure}
 \text{\hspace{1.0cm}} \resizebox{9cm}{!}{\includegraphics [angle=0,width=\textwidth] {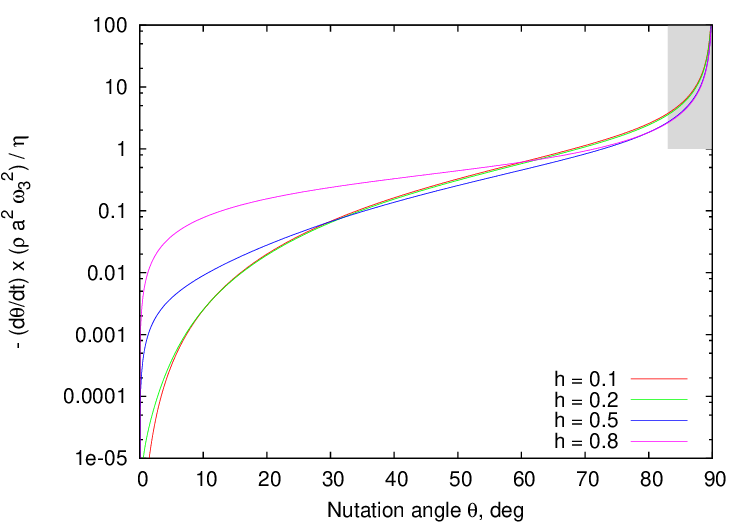}}
~\\
~\\
\caption{Normalised relaxation rate $\;({d \theta}/{dt})\;\rho\,a^2\,\omega_3^2\,\eta^{ \rm  -1}\,$ for a viscoelastic oblate ellipsoid (eqn. \ref{q}), for various values of the shape parameter $\,h=c/a\,$. Since $\,d\theta/dt\,$ is singular at $\,\theta = 90^\circ\,$, the plots are clipped at $\,\theta = 89.99^\circ\,$. The upper right ends of the curves (shaded) should be taken with caution, because in reality they cannot become asymptotically vertical in the limit of the initial $\theta\rightarrow 90^{ \rm  \circ}$. Indeed, in this limit the adiabaticity assumption (\ref{13}) no longer works and our description is no longer valid. See the text.}
\label{fig4}
\end{figure}

 The power $\,P\,$ (equations \ref{eq1} and \ref{eq3}) is zero for $\,\theta = 0\,$ and is finite for $\,\theta = \pi/2\,$ (see Figure \ref{fig3}). Naturally, for all the curves depicted in Figure \ref{fig3}, $\,P\,$ is the highest at $\,\theta = \pi/2\,$: \,the farther the rotator from the energy minimum the stronger its (so to say) ``desire'' to get to that minimum.$\,$\footnote{~It turns out, however, that for very high values of $\,h\,$ (for $\,h \gtrsim 0.95\,$ in the case of an oblate ellipsoid, and for $\,h \gtrsim 0.78\,$ in the case of a rectangular prism) $\,P\,$ attains its maximum at some values of $\,\theta\,$ slightly lower than $\,\pi/2\,$. This is a purely mathematical artefact called into being by the failure of the adiabaticity assumption (\ref{13}) in the limit of $\,h\rightarrow 1\,$, as well as in the limit of $\,\theta\rightarrow\pi/2\,$. This can be understood from the expression (\ref{omegafreq}) for $\,\omega\,$. Indeed, as can be seen from the subsequent expression (\ref{H}) for $\,H\,$, the precession frequency $\,\omega\,$ approaches zero as $\,h\rightarrow 1\,$. Also, as ensues from the expression (\ref{omegafreq}) for $\,\Omega_3\,$, the precession frequency $\,\omega\,$ approaches zero when $\,\theta\rightarrow\pi/2\,$. On both occasions, the vanishing of $\,\omega\,$ invalidates the adiabaticity assumption (\ref{13}).

 The maximal permissible value of the initial $\,\theta\,$ depends on the physical parameters of the rotator, including its viscosity $\eta$. Indeed, to find this limitation, we have to compare the precession rate $\,\omega\,$ given by formula (\ref{omegafreq}) and the precession relaxation rate given by formula (\ref{q}). The latter formula contains the power $\,P\,$ and, thereby, the viscosity and other parameters.
 \label{failure}}

 The plots depicted in Figure \ref{fig4} give a wrong impression that the relaxation rate $d\theta/dt$ diverges when the nutation angle $\theta$ is set to be $\pi/2$. In other words, these plots indicate a ``jump start'' of relaxation when the nutation angle is very large. In reality, while a start should indeed be faster for a larger initial $\theta$, an actual curve cannot become asymptotically vertical at $\theta\rightarrow\pi/2$. The reason for this is the same as in the case of the plot for the power, see Footnote \ref{failure}\,: ~our treatment is based on the adiabaticity assumption (\ref{13}) which no longer works in the upper right corner of the plot in Figure \ref{fig4}. So each of the shown curves must be subject to a cut-off at some value of the initial $\theta$ close to $\pi/2$. This is why we shaded the upper right corner in Figure \ref{fig4}.

 \subsection{Comparison with empirical theories based on an $\,${\it{ad hoc}}$\,$ quality factor}

 Basing their calculation on an empirically introduced quality factor $\,Q\,$, \citet{breiter+2012} arrived at the following expression for the relaxation time of an oblate spheroid:
 \ba
 T \;=\;-\;\frac{\mu\;Q}{a^2\;\rho\;\tilde{\omega}_3^3}\;\int_{ \rm  \theta_0}^{ \rm  \theta_f}\qquad\qquad\qquad\qquad\qquad\quad\quad
 \label{eq5}
 \ea
 \ba
 \nonumber
 \frac{
 (1+h^2)^4\;   d\theta
 }{
 \left(\frac{\textstyle 4}{\textstyle 7} \frac{\textstyle 25+20 h^2 + 16 h^4}{\textstyle 15 + 10 h^2 + 8 h^4}\,\sin^3
 \theta
 + \frac{\textstyle 8}{\textstyle 7} \frac{\textstyle 26+35 h^2}{\textstyle 13 + 20 h^2}\,h^4\,\sin \theta\,\cos^2 \theta \right)}
 \ea
 with $\,h\equiv c/a\,$ and $\,\tilde{\omega}_3\,\equiv\,J/I_3\,$.

 Earlier, \cite{efroimskylazarian2000} obtained the following approximation \footnote{~We call it approximation, because in \cite{efroimskylazarian2000} the boundary conditions on the sides of the prism were introduced approximately.} for the nutation relaxation time of a rotating rectangular prism:
 \begin{equation}
 T = \frac{-\;\mu\;Q }{a^2\,\rho\,\tilde{\omega}_3^3}  \int_{ \rm  \theta_0}^{ \rm  \theta_f}
 \frac{(1+h^2)^4\;d\theta }{ \frac{\textstyle 405}{\textstyle 108}\;\sin^3 \theta
 + \frac{\textstyle 5103}{\textstyle 432} h^4\,\sin \theta\;\cos^2 \theta}\;\;.
 \label{eq6}
 \end{equation}
 We depict these two results, (\ref{eq5}) and (\ref{eq6}), in Figure \ref{fig2}. The plots give the normalised dimensionless damping times $\,T\;\frac{\textstyle a^2 \rho~ \tilde{\omega}_3^3 }{\textstyle \mu\,Q}\,$ as functions of $\,h\,$, for different values of the initial nutation angle $\,\theta_0\,$ and the same final value $\,\theta_f\,$.

 One of the main differences between the damping times obtained through a consistent viscoelastic model and those obtained within the empirical $\,Q$-model is the presence of an additional factor $\,\cos \theta\,$ in the numerators of the viscoelastic-model expressions (\ref{eq2}) and (\ref{eq4}). The expressions (\ref{eq5}) and (\ref{eq6}) from the $\,Q$-model lack that factor in their numerators.$\,$\footnote{~Equivalently, we could have stated that the $\,Q$-model expressions contain an extra $\,\cos\theta\,$ in their denominators, as compared to the viscoelastic-model expressions.} The origin of this difference can be traced back to an extra factor $\,\cos \theta\,$ in the approximate expression for the power obtained within the $\,Q$-model. (Our formulae (\ref{eq1}) and (\ref{eq3}) lack that extra factor of $\,\cos\theta\,$.) \footnote{~To illustrate the emergence of a redundant $\,\cos \theta\,$ factor in the expression for power within the $\,Q$-model, consider a simplistic setting where
  the power is generated by a one-dimensional deviatoric stress at one frequency: $\,\sigma(t)\,=\,\sigma_0\,\cos\chi t\,$. The formulae (\ref{powerQ}) and (\ref{powermaxwell}) in the Appendix give the power calculated within the $\,Q$-model and the Maxwell model, correspondingly. In our case, the role of the forcing frequency $\,\chi\,$ in the equation (\ref{powerQ}) is played by the precession frequency $\,\omega\,$ defined by the expression (\ref{omegafreq}). The precession frequency depends on $\,\cos \theta\,$ through the definition of $\,\Omega_3\,$ in the equation (\ref{omegafreq}).}

 As the generated power computed with aid of the $\,Q$-model contains an additional factor $\,\cos \theta\,$, the power becomes zero in the limit of $\,\theta \rightarrow \pi/2\,$. This renders the $\,\theta-$dependence of the power quite different from that obtained in the previous section where we used the viscoelastic Maxwell model. The power generated in a viscoelastic oblate ellipsoid is shown in Figure \ref{fig3}. The power is maximal for $\,\theta = \pi/2\,$. In the case of very large values of the shape parameter ($\,h$ $\gtrsim 0.95\,$), the peak shifts to the values of $\,\theta\,$ slightly smaller than $\,\pi/2\,$~---~which is a purely mathematical artefact explained above in Footnote \ref{failure}.

 Within the viscoelastic model, the relaxation rate $\,d\theta / dt\,$ tends to infinity for $\,\theta  \rightarrow \pi/2\,$, see Figure \ref{fig4}.
 Within the empirical $\,Q$-model, the relaxation rate stays finite for $\,\theta  \rightarrow \pi/2\,$. It attains its maximum at an angle close to $\,\theta = \pi/2\,$, except for large $\,h\,$, in which case the peak shifts towards smaller values of $\,\theta\,$. (Such a shift resembles the $\,\theta-$dependence of the viscoelastic power, described in the previous section.). This explains the increasingly large differences between the curves with different initial $\,\theta_0\,$ for large and increasing $\,h\,$, seen in Figure \ref{fig2}.

 It can also be observed from Figure \ref{fig2} that within the empirical $\,Q$-model the damping times stay finite for spherical objects ($\,h \rightarrow 1\,$), which is obviously unphysical. At the same time, the consistent viscoelastic approach is spared of this defect. Owing to the factor $\,1-h^2\,$, the equations (\ref{eq2}) and (\ref{eq4}) furnish zero damping times for spherical objects. Our expressions are thus valid for the entire range of values of the shape parameter $\,h\,$.

 Both the generated power and the relaxation time resulting from our calculations turn out to depend solely on the viscosity $\,\eta\,$ of the Maxwell model.
 This is a consequence of using a particular rheological model, that of Maxwell. As demonstrated in Appendix \ref{AppendixE}, employment of a different rheological model (the one of Kelvin-Voigt) yields the power and the relaxation time dependent on both the viscosity $\,\eta\,$ and rigidity $\,\mu\,$.

 Similarly to \cite{sharma+2005} and \cite{breiter+2012}, in our calculations we accepted Poisson's ratio to be $\,\nu = 0.25\,$. This enabled us to use the elastic stress tensor computed by \cite{sharma+2005}. Such a choice simplifies the stress tensor considerably and renders it independent of any rheological parameters.$\,$\footnote{~In the general case (i.e., for an arbitrary value of Poisson's ratio), the distribution of stress depends on rheology due to the so-called compatibility conditions \citep[eqn. 10]{breiter+2012}.} This, in turn, significantly reduces the algebraic work needed to calculate the power. Similar analysis for an arbitrary value of Poisson's ratio should use a complete formulation for the stress tensor, provided in \cite{breiter+2012}.

 \subsection{Examples}

Consider a near-oblate homogeneous body having the size and density similar to those of the present-day Enceladus with $\,a\,\sim\,2.54\times 10^5$ m,
$\,\rho\,=\,1.61\times 10^3$ kg/m$^3$ \citep{thomas2016} and synchronised at the current Enceladean orbit ($\,\tilde{\omega}_3\,=\,n\,=\,5.31\times 10^{ \rm  -5}$ rad/s) .
Representative of an early undifferentiated Enceladus, this body will, for the purpose of our estimate, be treated as near-oblate, because its current triaxiality is $\,(I_2-I_1)/I_3\,=\,1.89\times 10^{ \rm  -2}\,$, this value ensuing from a model by \cite{thomas2016}.$\,$\footnote{~For a homogeneous ellipsoid, the dynamical triaxiality is equal to
$\;\frac{\textstyle I_2-I_1}{\textstyle I_3}\,=\,\frac{\textstyle a^2-\,b^2}{\textstyle a^2+\,b^2}\,\;,\;\,$
where the semiaxes, for Enceladus, are $\,a=256.2$ km, $\,b=251.4$ km, $\,c=248.6$ km. Insertion of these numbers in the said formula gives: $\,(I_2-I_1)/I_3\,=\,1.89\times 10^{ \rm  -2}\,$.} We also shall set the ratio of dimensions to be $\,c/a\,=\,0.97\,$, which once again is not very different from that of today's Enceladus.

While the present-day Enceladus should possess a viscosity value not very different from that of ice near melting point (likely, between $\,10^{ \rm  13}\,$ and $\,10^{ \rm  14}$ Pa s), we shall endow our object with a higher viscosity appropriate to a primitive icy body: $\,\eta\,=\,10^{ \rm  17}$ Pa s.

Now suppose that this ``early Enceladus'' was hit by a planetesimal and went into a wobble with the initial half-angle of the precession cone being $\,20^{ \rm  \circ}\,$. From equation (\ref{eq2}), we find that for this rotator the normalised dimensionless wobble-damping timescale assumes the value of about $\,0.3\;$:
  \ba
  T_{ \rm  dim.less}\,\equiv\;T\;\frac{\textstyle \rho a^2 \tilde{\omega}_3^2}{\textstyle \eta}\;\approx 0.3\,\;,
  \label{}
  \ea
  which entails:
  \ba
  T\;=\;\frac{\textstyle \eta}{\textstyle \rho a^2 \tilde{\omega}_3^2}\;T_{ \rm  dim.less}\;\approx\;3\times 10^3\;\mbox{yr}\;\,.
  \label{}
  \ea
As can be seen from the green and blue plots in Figure \ref{fig1}, the increase of the initial precession half-angle for this value of the $\,h\,$ will make the damping timescale longer by a factor of two, or so. On the other hand, if we increase the viscosity value even by two orders of magnitude, the resulting timescale will not even exceed 1 My, a relatively short time span in the life of a satellite. Thus we see that icy moons, after being pushed into a wobble by collisions, return quickly to a principal spin state.

 Also consider a hypothetical comet with a 10 km diameter, a rotation period of 10 hr and a density $\,\rho\,=\,0.6\times 10^3$ kg/m$^{ \rm  3}$ \citep{britt2006}. Typical for the known comets, this density is only slightly higher than that of a highly suppressed snow, the excess explainable by the presence of rock and ice.  The viscosity of snow varies between $\,10^8$ Pa s and $\,10^{ \rm  14}$ Pa s \citep{scapozza2004,stoffel2006}, with the upper limit obviously corresponding to a suppressed and dense snow. If we choose the said upper limit, we see that the nutation damping time is
  \ba
  \frac{\textstyle \eta}{\textstyle \rho a^2 \tilde{\omega}_3^2}\;\approx\;7\times 10^3\quad\Rightarrow\quad
  T\;\approx\;7\times 10^3\;T_{ \rm  dim.less}\;\,.
  \label{}
  \ea
  This renders $T \sim \,10^4-10^5$ yr for the oblateness coefficient $\,h\,$ between 0.5 and 0.9. For less dense comets, the viscosity should be lower and, accordingly, the damping time should be shorter. On the other hand, the damping times for asteroids may be much longer (possibly, up to Gyr), because their mean viscosity should exceed that of ice or snow by orders of magnitude.

 \subsection{On the applicability of the Maxwell model to asteroids}

 As a starting point of our project, we assumed that the unperturbed (no-wobble) shape of the asteroid is oblate. This form (sometimes referred to as a $\,${\it{Maclaurin ellipsoid}}$\,$) is an equilibrium shape of a non-precessing viscoelastic rotator. Our formalism can also be extended to a triaxial equilibrium shape called a $\,${\it{Jacobi ellipsoid}}$\,$ (such equilibrium shapes emerge at higher spin rates~---~see Footnote \ref{footnote4} in Section \ref{hi}).

 In Appendix \ref{solids} we explain the Maxwell model. Apart from its applicability to both ices and silicates at low frequencies, this rheological model has the advantage of mathematical simplicity.

 Many asteroids, however, are now believed to be rubble; so we should enquire how exactly the Maxwell model suites loosely connected matter in microgravity. Provided the model works, the next question should be on the acceptable values of the parameters entering it. While the effective rigidity of such materials can be estimated \citep{goldreichsari2009}, we still lack a handle on their effective viscosity.

 Justification of treating rubble as a viscoelastic material comes from numerical modeling ~---~ see, e.g., the work by Walsh, Richardson \& Michel
 (\citeyear{walsh+2008}) who studied both monodisperse (same-size) piles and simple bimodal distributions (those containing two different sizes of particles). It was found that monodisperse aggregates behave in a manner distinct from fluid, while mixing of spheres of different sizes yields a closer-to-fluid behaviour. Within a somewhat different model by \cite{tanga+2009}, even monodisperse piles demonstrated hydrodynamical behaviour. At the same time, simulated rubble piles did not behave like perfect fluids, due to their ability to sustain shear stress.$\,$\footnote{~Also see the experimental study by \cite{murdoch+2013}.} This means that the aggregates have a finite shear rigidity $\,\mu\,$ and may be regarded viscoelastic.

 With a cautious encouragement from the afore-cited works, we shall treat an asteroid as a Maxwell body, with intention to combine our algorithm with richer models when they appear. We expect that more advanced rheological laws for loose media may include effective rheological parameters other than the effective rigidity and viscosity.

 \section{Conclusions}

 We have calculated the nutation relaxation time for a viscoelastic oblate rotator.  The relaxation is caused by inelastic dissipation which, in its turn, results from the body's alternating deformation under the stressing provided by the inertial forces in the course of nutation.  The main novelty of our approach lies in that we described the inelastic response of the body by directly using its rheology, without resorting to an empirical quality factor $\,Q\,$.

 We started with calculation of the stress tensor in a nutating rotator, and then employed the rheological law and the so-called correspondence principle to obtain the strain tensor. While the correspondence principle is compatible with an arbitrary linear rheology, we chose the viscoelastic Maxwell model, because of its mathematical simplicity and applicability to both icy and silicate solids at low frequencies.
 %  $\,$\footnote{~There also exists evidence of rubble piles being viscoelastic, though more research in this direction is needed.}
 The so-calculated strain is naturally lagging in time behind the stress, which renders a non-zero dissipated power. Having computed this power, we found the relaxation rate and timescale.

 When the Maxwell model is employed, both the dissipated power and the nutation relaxation rate come out dependent on the shear viscosity $\,\eta\,$ only, not on the rigidity. Mathematically, $\,\eta\,$ assumes the role of the product $\,\mu\,Q\,$ that appeared in an old empirical model based on an {\it{ad hoc}} quality factor $\,Q\,$. However, even with $\,\eta\,$ inserted instead of $\,\mu\,Q\,$ into the old formulae, those differ from ours. In contradistinction from the preceding works, our approach renders the dissipated power vanishing for a shape approaching sphere (relaxation rate going to infinity). this not being the case for the old theories illustrated by Figure \ref{fig2}.

 Finally, it should be reiterated that our method is applicable to any arbitrary linear rheology.

\section*{Acknowledgements}

 \noindent
  We gladly thank Ond{\u{r}}ej \u{C}adek and David Vokrouhlick{\'{y}} for highly stimulating discussions which motivated us to look deeper into the problem. We also wish to thank Petr Pravec for a valuable consultation on the LCDB light curve database and for drawing our attention to the papers by \citet{pravec2005}, \citet{scheirich2010} and  \citet{pravec2014}.
  Our special gratitude goes to S{{\l}}awomir Breiter who refereed the manuscript for the Journal, and provided numerous comments that served to improve the quality of the paper.

%%%%%%%%%%%%%%%%%%%%%%%%%%%%%%%%%%%%%%%%%%%%%%%%%%
%%%%%%%%%%%%%%%%%%%% REFERENCES %%%%%%%%%%%%%%%%%%

%%%%%%%%%%%%%%%%%%%%%%%%%%%%%%%%%%%%%%%%%%%%%%%%%%
%%%%%%%%%%%%%%%%% APPENDICES %%%%%%%%%%%%%%%%%%%%%

 \appendix

 \section{The quality factor and the phase lag}\label{AppendixA}

 Consider a harmonic force
 \begin{equation}
 F = F_0 \cos \omega t
 \label{AAAAA3}
 \end{equation}
 acting on a system and causing a delayed displacement
 \begin{equation}
 x = x_0 \cos(\chi\,t - \epsilon)
 \label{AAAAA4}
 \end{equation}
 and the velocity
 \ba
 \nonumber
 v &=& -\;\chi\;x_0\;\sin(\chi t - \epsilon)\\
   &=& -\;\chi\;x_0 (\;\sin\chi t\;\cos\epsilon\;-\;\cos\chi t\;\sin\epsilon)\;.\qquad\qquad~\quad
 \label{472}
 \ea
 The resulting power is then equal to
 \begin{equation}
 {\cal P}\;=\;-\;A\;(\sin\chi t\;\cos\epsilon\;-\;\cos\chi t\;\sin\epsilon)\;\cos\chi t\;,
 \label{}
 \end{equation}
 with $\,A = F_0~\chi\;x_0\,$, whence the work performed over a time interval $\,(t_0\,,\;t)\,$ is:
 \be
 \nonumber
 w|^{ \rm   \textstyle{^{ \rm  ~t}}}_{ \rm  \textstyle{_{ \rm  ~t_0}}} = \int_{ \rm  t_0}^t{\cal P}~dt
 \ee
 \be
 \left.~\right.=\; A\int^{ \rm  \chi t}_{ \rm  \chi t_0}(\sin\chi t\,\cos\epsilon-\cos\chi t\,\sin\epsilon) \cos(\chi t)\,d(\chi\,t) \;\quad
 \ee
 \ba
 \nonumber
 &=& A\;\cos\epsilon~\int_{ \rm  \chi t_0}^{ \rm  \chi t}\cos z\;\sin z\;dz - A\;\sin\epsilon\int_{ \rm  \chi t_0}^{ \rm  \chi t}\cos^2 z\;dz ~\\
 \nonumber\\
 &=& - \frac{A}{4}~{\left[ \right.}\,\cos(2\chi t-\epsilon)\,+\,2\;\chi\;t\;\sin\epsilon~{\left.\right]}^{ \rm    \textstyle{^{ \rm  ~t}}}_{ \rm  \textstyle{_{ \rm  ~t_0}}}~~.
 \label{eqdissipation}
 \label{dissipation}
 \ea
 The oscillating term on the right-hand side gives the energy of deformation, stored in the system. The linear in time term renders the energy dissipated.$\,$\footnote{~This interpretation of the two terms was suggested by Stan Peale (2011, personal communication).}

 \noindent
 Over a cycle, the product $\,\chi t\,$ in the second term changes by $\,2\pi\,$, so the energy loss is
 \begin{equation}
 \Delta E_{ \rm  \textstyle{_{ \rm  cycle}}}\,=\;-\;A\;\pi\;\sin\epsilon
 % \;=\;-\;F_0~\chi\;x_0\;\pi\;\sin\epsilon
 ~.
 \label{ged}
 \end{equation}
 To calculate the peak energy accumulated in the system, note that the first term on the right-hand side of the equation (\ref{eqdissipation}) is
 maximal when integration is carried out over the interval from $\,\chi\,t\,=\,\pi/4\,+\,|\epsilon|/2\,$ through $\,\chi\,t\,=\,3\pi/4\,+\,|\epsilon|/2~$:
 \begin{equation}
 E_{ \rm  \textstyle{_{ \rm  peak}}}\,=\;~\frac{A}{2}~,
 \label{}
 \end{equation}
 Then the resulting inverse quality factor is:
 \begin{equation}
 Q^{ \rm  -1}\,=~\frac{~-~\Delta E_{ \rm  \textstyle{_{ \rm  cycle}}}}{2\,\pi\,E_{ \rm  \textstyle{_{ \rm  peak}}}}~=~
 \sin\epsilon~,\qquad
 \label{sin}
 \end{equation}
 not $\,\tan\epsilon\,$, $\,$as some believe.

 \section{Stresses and strains in a viscoelastic body}\label{AppendixC}

 Deformation of a precessing asteroid being extremely weak, the problem of precession relaxation can be assumed linear. This means that the current value of the strain tensor is linear in the values of the stress tensor.

 \subsection{Constitutive equation (rheological law)\label{rheology}}

 Were the medium perfectly elastic, the linearity law would be instantaneous. This means that the strain tensor at a certain instant of time, $\,e_{ \rm  \gamma\nu}(t)\,$, would be proportional to the stress tensor at that same instant, $\,\sigma_{ \rm  \gamma\nu}(t)\,$. In actual situations, linearity implies that the present strain $\,e_{ \rm  \gamma\nu}(t)\,$ contains a linear input from the present stress $\,\sigma_{ \rm  \gamma\nu}(t)\,$, as well as linear inputs from the past values of stress, $\,\sigma_{ \rm  \gamma\nu}(t\,')\,,$ with $\,t\,'<t\,$. Altogether, this gives a convolution operator:
 \begin{equation}
 2\,e_{ \rm  \gamma\nu}(t)\,=\,\hat{J}(t)~\sigma_{ \rm  \gamma\nu}\,=\,\int^{ \rm  t}_{ \rm  -\infty}\stackrel{\;\centerdot}{J}(t-t\,')~
 {\sigma}_{ \rm  \gamma\nu}(t\,')\,d t\,'
 \label{I12_4}
 \end{equation}
 often termed as the $\,${\it{constitutive equation}}$\,$ or as the $\,${\it{rheological law}}. Its kernel is a time derivative of a function $\,{J}(t-t\,')\,$ which is named the $\,${\it{compliance function}}$\,$ and which carries all information on the behaviour of the material under stressing.$\,$\footnote{~The kernel $\,\stackrel{\;\centerdot}{J}(t-t\,')\,$ of the integral operator $\,\hat{J}(t)\,$ is scalar in an isotropic medium only. More generally, the operator reads as $~2\,e_{ \rm  \alpha\beta}(t)\,=\,\int^{ \rm  t}_{ \rm  -\infty}
 \stackrel{\;\centerdot}{J}_{ \rm  \alpha\beta\gamma\nu}(t-t\,')~
 {\sigma}_{ \rm  \gamma\nu}(t\,')\,d t~$. Also be mindful that the convolution (\ref{I12_4}) is ignorant of the position, thus implying homogeneity of the body. In inhomogeneous media, the formalism becomes more elaborate. Finally, our rheological law is written for the deviatoric components of stresses and strains, and ignores the bulk components -- an approximation valid for incompressible deformation.}

 In the frequency domain, the convolution (\ref{I12_4}) assumes the form of a product:
 \begin{equation}
 2\;\bar{e}_{ \rm  \gamma\nu}(\chi)\,=\;\bar{J}(\chi)\;\bar{\sigma}_{ \rm  \gamma\nu}(\chi)\;\;.
 \label{LLJJKK}
 \end{equation}
 $\bar{e}_{ \rm  \gamma\nu}(\chi)\,$ and $\,\bar{\sigma}_{ \rm  \gamma\nu}(\chi)\,$ being the Fourier images of strain and stress, and $\,\chi\,$ being the forcing
 frequency.
 The complex compliance $\,\bar{J}(\chi)\,$ is the Fourier image of the kernel $\,\dot{J}(t-t\,')\,$ of the integral operator (\ref{I12_4})~---~see, e.g., \cite{efroimsky2012a,efroimsky2012b}.

 We see that in a linear regime the stresses evolving at some frequency give birth to strains at that same frequency, and cause no influence upon deformation at other frequencies. Simply speaking, different frequencies ``do not talk" to one another.

 Parameterising the reaction of a small sample of the material, the constitutive equation does not fully define the response of a celestial body consisting of this material. The response of the body is determined also by self-gravitation. This is, for example, the reason why the frequency-dependence of a tidal quality factor deviates from that of the seismic quality factor \citep{efroimsky2012a,efroimsky2012b}. In our treatment of the precession relaxation problem, we shall encounter a similar situation.

 \subsection{Solids. The Maxwell model}\label{solids}

  Over low frequencies, deformation of solids is viscoelastic and obeys the Maxwell model. At higher frequencies, more complicated processes enter the picture (mainly, unjamming of defects), and the response is more adequately described by the Andrade model. The borderline between the frequencies deemed low and high is determined by the temperature of the body (see \cite{efroimsky2012b} and references therein).

 Limiting our description to the Maxwell model, we shall assume that a deviatoric stress $\,{\boldsymbol{\sigma}}\,$ produces a deviatoric strain consisting of two parts: a purely elastic part $\,\stackrel{(e)}{\boldsymbol{e}}\,$ and a purely viscous part $\,\stackrel{(v)}{\boldsymbol{e}}\;$:
     \begin{equation}
     \begin{split}
       {\boldsymbol{e}}\,=\;\stackrel{(e)}{\boldsymbol{e}}\,+\,\stackrel{(v)}{\boldsymbol{e}}  ~,\,\quad & \mbox{where}\,\quad {\boldsymbol{\sigma}}\,=\,2\,\mu\,\stackrel{(e)}{\boldsymbol{e}}~~~~ \\
      & \mbox{and}\quad~ {\boldsymbol{\sigma}}\,=\,2\,\eta\,\frac{\partial\,}{\partial t}~\stackrel{(v)}{\boldsymbol{e}}  ~,
      \label{dddt}
     \end{split}
    \end{equation}
 $\eta\,$ and $\,\mu\,$ being the viscosity and unrelaxed rigidity. These formulae, together, give us:
 \begin{subequations}
  \begin{equation}
 \stackrel{\centerdot}{\boldsymbol{e}}\,=\,\frac{1}{2\,\mu}\;\stackrel{\centerdot}{\boldsymbol{\sigma}}\,+\,\frac{1}{2\,\eta}\;{\boldsymbol{\sigma}}
 \label{}
  \end{equation}
 or, in an equivalent form:
 \begin{equation}
 \stackrel{\centerdot}{\boldsymbol{\sigma}}\,+\;\frac{1\;}{\tau_{ \rm  _M}}\,{\boldsymbol{\sigma}}
 ~=~2\,\mu\,\stackrel{\centerdot}{\boldsymbol{e}}~,
 \label{}
 \end{equation}
 \label{these}
 \end{subequations}
 where $\,\tau_{ \rm  _M}\,$ is the {\emph{Maxwell time}} defined as
 \begin{equation}
 \tau_{ \rm  _M}\;\equiv\;\frac{\,\eta\,}{\,\mu\,}~.
 \label{Maxwell}
 \end{equation}
 In the case of bulk deformation, we would have to use the bulk parameters instead of their deviatoric counterparts: $\,3K\,$ instead of $\,2\mu\,$, and $\,\zeta\,$ instead of $\,\eta\,$.

 The equation (\ref{these}) can be cast into the integral form (\ref{I12_4}), with the kernel being a time derivative of the compliance function
  \begin{equation}
 ^{ \rm  \textstyle{^{ \rm  (Maxwell)}}}J(t\,-\,t\,')\,=\,\left[\,J\,+\,\left(t\;-\;t\,'\right)\;\frac{1}{\eta}\,\right]\;\Theta(t\,-\,t\,')~,
 \label{Max}
  \end{equation}
 $\Theta(t\,-\,t\,')\,$ being the Heaviside step function, and $\,J\,$ being the unrelaxed compliance:
  \begin{equation}
 J~\equiv~\frac{\,1\,}{\,\mu\,}~.
 \label{}
  \end{equation}
 In the frequency domain, the equation (\ref{these}) can be written in the form of (\ref{LLJJKK}), with the complex
 %  rigidity and
 compliance set to be
 \begin{equation}
 ^{ \rm  \textstyle{^{ \rm  (Maxwell)}}}{\bar{\mathit{J\,}}}(\chi)~=~J\,-\,\frac{i}{\eta\chi}~=~J\,\left(\,1~-~\frac{i}{\chi\,\tau_{ \rm  _M}}\right)~,
 \label{don}
 \end{equation}
 the terms $\,J\,$ and $\,-\,{i}/(\eta\chi)\,$ giving the elastic and viscous parts of deformation, correspondingly. Such a body becomes elastic at high and viscous at low frequencies.

 The ``seismic" phase lag $\,\delta\,$ (as distinct from the tidal phase lag) is defined through
  \begin{equation}
 \bar{J}(\chi)~=~|\,\bar{J}(\chi)\,|~\exp[\,-\,i\,\delta(\chi)\,]~,
 \label{}
  \end{equation}
  with a ``minus" sign needed to endow $\,\delta\,$ with the meaning of a delay of the reaction (strain) with respect to the action (stress).

 In the case of the Maxwell model, the imaginary part and the absolute value of the compliance are given by
  \begin{equation}
 % {\cal{I}}{\it{m}}
 \Im \left[\,\bar{J}(\chi)\,\right]\,=\,-\,|\,\bar{J}(\chi)\,|\,\sin\delta(\chi)\,=
 \,-\,\frac{J}{\chi\,\tau_{ \rm  _M}}\,=\,-\,\frac{1}{\chi\,\eta}
 \label{anaconda}
  \end{equation}
  and
  \be
  |\,\bar{J}(\chi)\,|\;=\;\sqrt{J^2\,+\;\frac{\textstyle{1}}{\textstyle{(\chi\,\eta)^2}}}~,
  \ee
 wherefrom
  \begin{equation}
 \sin\delta(\chi)~=~-~\frac{1}{\sqrt{\chi^2\,\tau_{ \rm  _M}^2\,+\,1}}\quad.
 \label{angle}
  \end{equation}

 Mind that here the rigidity and compliance assume their $\,${\it{unrelaxed}}$\,$ values:
  \begin{equation}
 \mu~=~\mu(0)\qquad\mbox{and}\qquad J~=~J(0)~.
 \label{}
  \end{equation}

 \subsection{A finite-size Maxwell sphere versus a small Maxwell sample}

 In a periodically deformed sample of a Maxwell material, the strain will lag behind the applied stress. According to the formula (\ref{don}), the phase lag $\,\delta\,$ satisfies
 \begin{equation}
 \tan \delta(\chi)~=~-~\frac{\Im[\bar{J}(\chi)]}{\Re[\bar{J}(\chi)]}~=~(\tau_{ \rm  _M}\, \chi)^{ \rm  -1}~.
 \label{}
 \end{equation}
 Very importantly, this expression will $\,${\it{not}}$\,$ hold for the lag in a celestial body composed of a Maxwell material. The response of a celestial body of a certain rheology differs from the response of a sample of this rheology. The two reasons for this are: (a) self-gravitation of the body, and (b) the inertial forces due to the body's rotation, proper or improper \citep{efroimsky2012a}.

 \section{Difficulties}\label{difficulties}

 Back in Section \ref{generic}, we raised two uneasy questions.
 \begin{itemize}
 \item[\bf 1.~] Should we take into account the oscillating part $\,\tilde{\bf{b}}_{ \rm  gr}\,$ of the gravity force, generated by the periodically changing deformation caused by the inertial forces?
 \item[\bf 2.~] Should we keep the constant prestressing that comprises the constant part of the gravity field and the constant part of the centrifugal force?
 \end{itemize}
 Here we shall consider a simplified version of our problem.

 \subsection{Simplify the problem: ~drop the toroidal force}

 As well known \cite[see, e.g.,][$\,$Appendices B3 - B4]{efroimsky2012a}, the inertial forces can be split into a toroidal part, a radial part, and a potential part of degree $\,l=2\,$. The radial part is, mathematically, equivalent to a decrease in the Newton gravitational constant, and thus can be omitted (effectively, absorbed by self-gravitation). Then, in neglect of the toroidal input, our problem can be reduced to a simplified version equivalent to degree-2 tides. This makes mathematics much easier, nevertheless granting us an opportunity to find answers to the above questions.

 Static tides obey a boundary-value problem for a system of two equations:
  \begin{eqnarray}
 \sigma_{ \rm  \textstyle{_{ \rm  \beta\nu}}}&=&2\;\mu\;e_{ \rm  \textstyle{_{ \rm  \beta\nu}}}~, \label{tide1}\\
 \nonumber\\
 0&=&\frac{\partial \sigma_{ \rm  \textstyle{_{ \rm  \beta\nu}}}}{\partial x_{ \rm  \textstyle{_\nu}}}\;-\;\rho\;\frac{\partial (W_2\,+\,U_2)}{\partial
 x_{ \rm  \textstyle{_\beta}}}~.
 \label{tide2}
  \end{eqnarray}
 One equation interconnects the static stress and strain tensors via the relaxed shear rigidity modulus $\,\mu\,$. The other is the Second Law of Newton. The notations $\,\sigma_{ \rm  \textstyle{_{ \rm  \beta\nu}}}\,$ and $\,e_{ \rm  \textstyle{_{ \rm  \beta\nu}}}\,$ stand for the deviatoric stress and strain (the volumetric parts being neglected for simplicity). The quantity $\,W_2\,$ signifies the quadrupole perturbing potential. In our case, as demonstrated, e.g., in \cite{efroimsky2012a},
  \begin{equation}
 W_2(\bold{r})~=~\frac{\rho}{3}~ \bold{\Omega}^2\,\bold{r}^2\,P_2(\sin\phi)~,
 \label{W2}
  \end{equation}
 with $\,P_2(\sin\phi)\,=\,\frac{\textstyle 1}{\textstyle 2}\,(3\,\sin^2\phi\,-\,1)\,$ being a Legendre polynomial, and $\,\phi\,$ being the angle between the angular-velocity and position vectors:
    \begin{equation}
 \cos\phi\,~=~\frac{\bold{\Omega}}{\,|\bold{\Omega}|\,}\cdot\frac{\bold{r}}{\,|\bold{r}|\,}~.
    \label{cps} \end{equation}
  In the limit of oblate symmetry ($I_1=\,I_2$), this expression becomes (see the equation \ref{eq}):
 \ba
 \nonumber
 \cos\phi~  =~\frac{1}{\,|\bold{\Omega}|\;|\bold{r}|\,}\;\left(\;\pm\;x\;\Omega_{ \rm  \perp}\;\cos\omega (t-t_0)  \right.\quad\qquad~\qquad\\
 \left. +\;y\;\Omega_{ \rm  \perp}\;\sin\omega (t-t_0)\;+\;z\;\Omega_3
 \,\right)~,
 \label{coss}
 \ea
 where $\,\omega\,$ is the precession frequency (\ref{omegafreq}).

 The quantity $\,U_2\,$ is the additional tidal potential due to small variations of shape of the tidally perturbed body. Solution of the system (\ref{tide1} - \ref{tide2}) should give: $\,U_2\,=\,k_2\,W_2\,$, where $\,k_2\,$ is the quadrupole static Love number.

 For tides evolving in time, the system preserves its form under certain conditions,$\,$\footnote{~For a short explanation, see Appendix \ref{AppendixE} below. For a more detailed discussion, see Efroimsky (2012$\,$a, Appendix B).}
  with all entities becoming Fourier images: $\,\sigma_{ \rm  \textstyle{_{ \rm  \beta\nu}}}\rightarrow\bar{\sigma}_{ \rm  \textstyle{_{ \rm  \beta\nu}}}(\chi)\,$, $\,e_{ \rm  \textstyle{_{ \rm  \beta\nu}}}\rightarrow\bar{u}_{ \rm  \textstyle{_{ \rm  \beta\nu}}}(\chi)\,$, $\,\mu\rightarrow\bar{\mu}(\chi)\,$, $\,W\rightarrow\bar{W}(\chi)\,$, $\,U\rightarrow\bar{U}(\chi)\,$, where $\,\chi\,$ is the tidal frequency:
  \begin{eqnarray}
 \bar{\sigma}_{ \rm  \textstyle{_{ \rm  \beta\nu}}}(\chi)&=&2\;\bar{\mu}(\chi)\;\bar{u}_{ \rm  \textstyle{_{ \rm  \beta\nu}}}(\chi)~~~,
 \label{tide3}\\
 \nonumber\\
 0&=&\frac{\partial \bar{\sigma}_{ \rm  \textstyle{_{ \rm  \beta\nu}}}(\chi)}{\partial x_{ \rm  \textstyle{_\nu}}}\;-\;\rho\;\frac{\partial (\,\bar{W}_2(\chi)\,+\,\bar{U}_2(\chi)\,)}{\partial
 x_{ \rm  \textstyle{_\beta}}}~.\quad~\quad
 \label{tide4}
  \end{eqnarray}
 By solving the system, one must arrive at  $\,\bar{U}_2(\chi)=\bar{k}_2(\chi)\,\bar{W}_2(\chi)\,$, with $\,\bar{k}_2\,$ being the degree-$2\,$ dynamical Love numbers. Evidently, this goal would not be achieved, had the deformation-caused potential $\,U_2\,$ been neglected.

  We cannot neglect $\,U_2\,$ in our calculation of the picture of tidal stress either -- even if we seek only an approximate solution. Indeed, by dropping $\,U_2\,$ in the equation (\ref{tide4}), we would have arrived at a conclusion that the stress is in phase with the tide-raising potential $\,W_2\,$. This, however, would contradict
 the equation (\ref{tide3}) where lagging is mandatory and is prescribed by the rheological model $\,\bar{\mu}(\chi)=1/\bar{J}(\chi)\,$ -- ~see Appendix \ref{AppendixC}.

 Now, regarding the dissipation rate. At first glance, this is the case where the response $\,U_2\,$ can be omitted. In reality, this turns out to be a nontrivial matter.

 \subsection{Dissipation rate. Calculation through velocities and forces}\label{forces}

 The dissipated power is
  \begin{equation}
 P~=~\int dV~\rho~{\bf\dot{u}}\cdot{\bf{F}}~=~-~\int dV~\rho~{\dot{u}}_\alpha\,\left(\,\frac{\partial W_2}{\partial x_\alpha}~+~\frac{\partial U_2}{\partial x_\alpha}\,\right)~,
 \label{}
   \end{equation}
 where the response potential $\,U_2\,$ is in phase with the deformation $\,{\bf{u}}\,$ and, thereby, is in quadrature with the velocity $\,{\bf\dot{u}}\,$. So, after time-averaging, the $\,\partial_{ \rm  \alpha}U_2\,$ term drops out:
   \begin{equation}
 P
 ~=~
 -~\int dV~\rho~{\dot{u}}_\alpha\,\frac{\partial W}{\partial x_\alpha}\,
 ~,
 \label{}
   \end{equation}
 where we omitted the time-average symbol $\,\langle\,...\rangle\,$.
 Integration by parts then leads us to
   \begin{equation}
   \begin{split}
 P~ & =~-~\int dV~\rho~{\dot{u}}_\alpha\,\frac{\partial W_2}{\partial x_\alpha}\; \\
 & = \;-\;\int dV~\partial_{ \rm  \alpha}\left(\,\rho\,W_2\,\dot{u}_\alpha\right)~+~\int dV~\rho~W_2~\partial_\alpha\,\dot{u}_\alpha~.~\quad
 \label{}
  \end{split}
   \end{equation}
 For incompressible media, the second term vanishes, whence
   \begin{equation}
      \begin{split}
 P~=\;\int dV~\partial_{ \rm  \alpha}\left(\,\rho\,W_2\,\dot{u}_\alpha\right)~ & =~-\int\rho~W_2~\dot{u}_\alpha\,n_{ \rm  \alpha}\,dS~ \\
  & \approx~-~\int\rho~W_2~\dot{u}_r\,dS
 ~,
 \label{tibet}
   \end{split}
   \end{equation}
 with $\,n_\alpha\,$ signifying the surface normal, and $\,\dot{u}_r\,$ being the radial deformation velocity near the surface.$\,$\footnote{~It takes effort to justify the approximation $\,\dot{u}_\alpha\,n_{ \rm  \alpha}\,\approx\,\dot{u}_r\,$, ~see \cite{efroimskymakarov2014}.} $\,$In the quadrupole approximation,
 the static elevation $\,u_r\,$ is
 \begin{subequations}
   \begin{equation}
 u_r\;=\;h_2\;\frac{W_2}{\mbox{g}}~,
 \label{}
   \end{equation}
 while for evolving tide a similar relation holds between the Fourier images:
   \begin{equation}
 \bar{u}_r(\chi)~=~\bar{h}_2(\chi)\;\frac{\bar{W}_2(\chi)}{\mbox{g}}~,
 \label{}
   \end{equation}
 \label{}
  \end{subequations}
 g being the surface gravity, and $\,\chi\,$ being the forcing frequency. Therefrom, the energy dissipation rate can be written as
 \begin{subequations}
   \begin{equation}
 P\;=\;-\;\int \frac{\rho}{\mbox{g}}~\sum_{ \rm  \chi}\bar{W}_2(\chi)~\chi~\bar{W}_2(\chi)~\bar{h}_2(\chi)~,
 \label{}
   \end{equation}
 with an extra factor of $\,\chi\,$ showing up due to the presence of time derivative in (\ref{tibet}), and with time averaging implied on both sides. Equivalently, this can also be written as
   \begin{eqnarray}
 P&=&-\;\int dS~\frac{\,5\,\rho\,}{\mbox{3\,g}}~\sum_{ \rm  \chi}\bar{W}_2(\chi)~\chi~\bar{W}_2(\chi)~\bar{k}_2(\chi)
 \label{}\\
  &=&-\;\int dS~\frac{\,5\,\rho\,}{3\,\mbox{g}}~\sum_{ \rm  \chi}\,\chi~\bar{W}_2(\chi)~\bar{U}_2(\chi)~.~\qquad~\qquad
 \label{pamir}
   \end{eqnarray}
 \label{}
  \end{subequations}
 Recalling that both the left- and right-hand sides are implied to be time-averaged, we see that the mean power is proportional to the sine of the tidal phase lag:
   \begin{equation}
 P~=~-~\int dS~\frac{\,5\,\rho\,}{3\,\mbox{g}}~\sum_{ \rm  \chi}\,\chi~|\,\bar{W}_2(\chi)\,|^2\,{k}_2(\chi)~\sin\epsilon_2(\chi)~,
 \label{power1}
   \end{equation}
 where the dynamical Love number is $\,{k}_2(\chi)\,\equiv\,|\,\bar{k}_2(\chi)\,|\,$.

 From this, two conclusions ensue.
 \begin{itemize}
 \item[1.~~~] The presence of the deformation-caused additional tidal potential $\,U_2\,$ in the expression (\ref{pamir}) tells us that the small deformation -- and the associated gravity variation -- cannot be dropped. On the contrary, they play a key role.\vspace{2.5mm}
 \item[2.~~~] The presence of the surface gravity g and the Love number $\,h_2\,$ (or of its kin $\,k_2\,$) indicates that the constant prestressing matters and cannot be dropped either.
 \end{itemize}

 In the subsequent section, we shall see that the former of these two difficulties can be sidestepped, while the latter can be neglected only for sufficiently high value of the product of the wobble frequency $\,\chi\,$ by the effective viscosity $\,\eta\,$.

 Finally, in the limit of vanishing wobble frequency $\,\chi\,$, the product $\,k_2(\chi)\,\sin\epsilon_2(\chi)\,$ becomes linear in $\,\chi\,$
 \citep{efroimsky2012a,efroimsky2012b}.
 As can be understood from the formulae (\ref{W2} - \ref{coss}), the Fourier component $\,W_2(\chi)\,$ stays finite in the limit of vanishing $\,\chi\,$. So the dissipated power (\ref{power1}) scales as $\,\chi^2\,$ in the said limit.

 At the same time, at higher frequencies, $\,k_2(\chi)\,\sin\epsilon_2(\chi)\,$ of a Maxwell sphere behaves as $\,\chi^{ \rm  -1}\,$ \citep{efroimsky2012a,efroimsky2012b}. Thence at higher values of $\,\chi\,$ the damping rate (\ref{power1}) is $\,\chi$-independent. These observations will help us in the next section.

 \subsection{Dissipation rate. Calculation via the strain rate and stress}\label{AppendixD3}

 Now let us see if the above complications can be sidestepped by switching from velocities and forces to strain rates and stresses:
   \begin{equation}
 P~=~\int dV~\rho~{\bf\dot{u}}\cdot{\bf{F}}~=~-~\int dV\,\rho\;{\dot{u}}_\alpha\;\partial_\beta\sigma_{ \rm  \alpha\beta}~.
 \label{}
   \end{equation}
 Integration by parts furnishes us with
 \begin{subequations}
   \begin{eqnarray}
 P&=&-~\int dV\;\partial_\beta\left(\rho\;{\dot{u}}_\alpha\;\sigma_{ \rm  \alpha\beta}\right)~+~
 \int dV\,\rho~\sigma_{ \rm  \alpha\beta}\,\partial_\beta {\dot{u}}_\alpha
 ~.\qquad
 \label{}\\
 %  \ea \ba
  &=&-~\int\rho\;{\dot{u}}_\alpha\;\sigma_{ \rm  \alpha\beta}\;n_{ \rm  \beta}~dS~+~
 \int dV\,\rho~\sigma_{ \rm  \alpha\beta}\,\partial_\beta {\dot{e}}_{ \rm  \alpha\beta}~,\qquad
 \label{}
   \end{eqnarray}
 \label{}
  \end{subequations}
 ${{e}}_{ \rm  \alpha\beta}\,$ being the strain tensor. Owing to the boundary conditions, the surface term vanishes, leaving us with
 \footnote{~In the expression (\ref{heineken}), the stress (containing both a constant and vibrating components) is multiplied by the oscillating strain velocity. After time averaging, the product of the constant part of the stress by the oscillating strain velocity gives zero. This is why in (\ref{pilzner}) only the oscillating part of the stress is kept.\vspace{1mm}\\
 The factor of $\,\chi\,$ emerges in (\ref{pilzner}) due to differentiation of the strain with respect to time.
  An extra factor of 1/2 shows up because the complex compliance is conventionally introduced as  $\,2\,\bar{e}_{ \rm  \alpha\beta}(\chi)\,=\,\bar{J}(\chi)~\bar{\sigma}_{ \rm  \alpha\beta}(\chi)\,$.\label{footnote}}
  \begin{subequations}
   \begin{eqnarray}
 P&=&\int dV\,\rho~\sigma_{ \rm  \alpha\beta}\,\partial_\beta {\dot{e}}_{ \rm  \alpha\beta}
 \label{heineken}\\
  &=&\int
 dV~\rho~\frac{1}{2}~\sum_{ \rm  \chi}\bar{\sigma}_{ \rm  \alpha\beta}(\chi)~\chi~\bar{J}(\chi)~\bar{\sigma}_{ \rm  \alpha\beta}(\chi)~.\qquad~\qquad
 \label{pilzner}
   \end{eqnarray}
 \label{}
 \end{subequations}
 After time averaging, we obtain:
   \begin{equation}
 P~=~-~\int dV~\rho~\frac{1}{2}~\sum_{ \rm  \chi}\,|\,\bar{\sigma}_{ \rm  \alpha\beta}(\chi)\,|^2~\chi~|\,\bar{J}(\chi)\,|~\sin\delta~,
 \label{power2}
   \end{equation}
 where $\,\delta\,$ is the ``seismic" (not tidal) lag defined through
   \begin{equation}
 \bar{J}(\chi)~=~|\,\bar{J}(\chi)\,|~\exp[\,-\,i\,\delta(\chi)\,]~.
 \label{}
   \end{equation}
 Specifically, for the Maxwell model, this lag is furnished by the equation (\ref{angle}).

 The neglect of small oscillations of shape does not nullify $\,|\,\bar{J}(\chi)\,|\,$, because this quantity is local and is defined by the rheology, not by the shape. Hence, within this formulation, it is permissible to ignore the small changes of the shape and the resulting oscillating component $\,\tilde{\bf{b}}_{ \rm  gr}\,$ of the gravity force.

 Now recall that the gradient of the stress is equal to the reaction force. The latter, in its turn, is equal to the body
 force,$\,$\footnote{~As we are working in the body frame, the overall body force incorporates the inertial forces~---~see the equation (\ref{28b}).}
 the evolving component whereof is proportional to the wobble frequency $\,\chi\,$:
   \begin{equation}
 \sigma_{ \rm  ij}\;\propto\;\chi~.
 \label{}
   \end{equation}
  This observation, along with the expression (\ref{anaconda}), tells us that the dissipated power is proportional to $\,\chi^2\,$. This coincides with the observation made in the end of the preceding subsection. Mind, though, that the said observation was valid for low frequencies only, while at high frequencies the power was demonstrated to be frequency-independent. To comply with this caveat, we have to adjust the above formula for the stress. We expect it to acquire the form of
   \begin{equation}
 \sigma_{ \rm  ij}\;\propto\;\frac{1}{\sqrt{(J\,+\,{\cal{B}}_{ \rm  \textstyle{_2}})^2\;+\;1/(\eta\,\chi)^2}}~,
 \label{guess}
   \end{equation}
 where
   \begin{equation}
 {\cal{B}}_{ \rm  2}\,\equiv~\frac{\textstyle{19}}{\textstyle{2}\,\mbox{g}\,
 \rho\,R}~=\;\frac{\textstyle{57}}{\textstyle{8\,\pi\,
 G\,\rho^2\,R^2}}~
 ,\qquad
 \label{B}
   \end{equation}
 $\rho\,$ and $\,R\,$ being the sphere's density and radius, and $\,G\,$ being the Newton gravitational constant.
 We must arrive at such an expression for the stress, if we correctly introduce the compatibility conditions. These conditions are merely a way to take care of prestressing and the boundary conditions.

 The formula (\ref{guess}) was an easy guess, since we know from Efroimsky (2012$\,$a,b) that
   \begin{equation}
 \bar{k}_{ \rm  2}(\chi)\;\sin\epsilon_{ \rm  2}(\chi)~=~-~\frac{3}{2}~\frac{{\cal{B}}_{ \rm  \textstyle{_2}}\;{\cal{I}}{\it{m}}\left[\bar{J}(\chi)\right]
 }{\left({\cal{R}}{\it{e}}\left[\bar{J}(\chi)\right]+{\cal{B}}_{ \rm  \textstyle{_2}}\right)^2+\left({\cal{I}}{\it{m}}
 \left[\bar{J}(\chi)\right]\right)^2}
 \label{E10a}
   \end{equation}
 In the denominator of this expression, the real and imaginary parts of $\,\bar{J}(\chi)\,$ show up due to self-gravitation. \footnote{~Trace the transition from the equation (36) to (40a) in Efroimsky (2015).
 }

 The expression (\ref{E10a}) also gives us a clue to understanding what ``low frequencies" and ``high frequencies" mean. For a spherical Maxwell body, the maximum of that function is located at
   \begin{equation}
 \chi_{ \rm  peak}\,=\;\frac{1}{\tau_{ \rm  _M}\,{\cal{B}}_{ \rm  l}\;\mu}\;=\;\frac{\textstyle{8\,\pi\,
 G\,\rho^2\,R^2}}{\textstyle{57\;\eta}}~.
 \label{}
   \end{equation}

 At wobble frequencies $\,\chi\,$ lower than $\,\chi_{ \rm  peak}\,$, prestressing matters. The product $\,\bar{k}_{ \rm  2}(\chi)\,\sin\epsilon_{ \rm  2}(\chi)\,$ is linear in $\,\chi\,$, and the expression (\ref{power1}) for $\,P\,$ scales as $\,\chi^2\,$. The expression (\ref{guess}) for the stress is linear in $\,\chi\,$, and the expression (\ref{power2}) for $\,P\,$ behaves as $\,\chi^2\,$ ~---~ which agrees with the low-frequency behaviour of the expression (\ref{power1}).

 At frequencies $\,\chi\,$ higher than $\,\chi_{ \rm  peak}\,$, prestressing plays no role. The $\,\bar{k}_{ \rm  2}(\chi)\;\sin\epsilon_{ \rm  2}(\chi)\,$ scales as $\,\chi^{ \rm  -1}\,$, and the expression (\ref{power1}) for $\,P\,$ becomes $\,\chi$-independent. The expression (\ref{guess}) for the stress is $\,\chi$-independent, and so is the expression (\ref{power2}) for $\,P\,$ ~---~ which agrees with the high-frequency behaviour of the expression (\ref{power1}).

 The condition of prestressing being unimportant,
   \begin{equation}
 \chi\;\gg\;\chi_{ \rm  peak}\;=\;\frac{\textstyle{8\,\pi\,
 G\,\rho^2\,R^2}}{\textstyle{57\;\eta}}~,
 \label{}
   \end{equation}
 is equivalent to
   \begin{equation}
 \eta\;\gg\;\frac{\textstyle{8\,\pi\,G\,\rho^2\,R^2}}{\textstyle{57\;\chi}}~.
 \label{109}
   \end{equation}
 So, for asteroids of viscosity higher than this threshold, prestressing may be ignored.

 This calculation was performed for a one-frequency case. In realistic situations, $\,\chi\,$ in the above criterion should be identified with the lowest forcing frequency of the deformation spectrum (in the case of a dynamically oblate rotator, with the precession frequency $\,\omega\,$).

 The condition (\ref{109}) of prestressing being irrelevant has been derived here in neglect of the toroidal part of the wobble-caused deformation. This is a touchy point of our treatment, wherefore the condition (\ref{109}) remains, to some extent, an article of faith. We believe that this estimate will not change much when the toroidal part of the perturbation is included into the picture~---~though a more rigorous analysis of this situation remains desirable.

 \section{The correspondence principle}\label{AppendixE}

 In continuum mechanics, there exists an important theorem called $\,${\it{elastic-viscoelastic analogy}}$\,$ (also referred to as  $\,${\it{correspondence principle}}). It establishes a mathematical identity between the equations describing a viscoelastic problem for evolving forces, strain and stress, on the one hand, and an elastic problem for evolving forces, strain and stress, on the other hand. Since the latter problem is mathematically equivalent to a static one, we may as well say that the theorem maps a viscoelastic problem with evolving deformation onto a corresponding static problem.

 While \citet{biot} is often credited with the discovery of this analogy, its early version was in fact suggested by \citet{darwin}.

 In short, the equation of motion and the constitutive equation describing a viscoelastic problem in the frequency domain are algebraically
 identical to the elastic (or static) counterparts of these equations. To obtain the viscoelastic equations from the elastic (or static) ones, the time-domain values of forces, strain and stress in the elastic problem (or the constant values in the static problem) should be substituted with the Fourier or Laplace images of the forces, strain and stress in the corresponding viscoelastic setting. At the same time, the elastic (or static) moduli and compliance should be substituted with their complex counterparts.
 \ba  \nonumber
   \left(
   \begin{array}{cc}
      \multicolumn{2}{c}{\textbf{Elastic}} \\~\\
      \sigma_{ \rm  ij,\,j}(\textbf{x})  \;= &  -\;\rho\;f_i(\textbf{x}) \\~\\
      \sigma_{ \rm  ij}^{ \rm  (D)}(\textbf{x}) \;= &  2\;\mu\;{e}_{ \rm  ij}^{ \rm  (D)}(\textbf{x})  \\~\\
      \sigma_{ \rm  ij}^{ \rm  (V)}(\textbf{x}) \;= &  3\;K\;{e}_{ \rm  ij}^{ \rm  (V)}(\textbf{x})   \\~\\
   \end{array}
   \right )
 \longrightarrow\\
 \label{}\\
\nonumber\\
\nonumber
   \left (
   \begin{array}{cc}
         \multicolumn{2}{c}{\textbf{Viscoelastic}} \\~\\
     \bar{\sigma}_{ \rm  ij,\,j}(\textbf{x},\,\chi) 	\;=	&   - \rho\;\bar{f}_i(\textbf{x},\,\chi)  \\~\\
      \bar{\sigma}_{ \rm  ij}^{ \rm  (D)}(\textbf{x},\,\chi)	\;=	&  2\; \bar{\mu}(\chi)\;\bar{{e}}_{ \rm  ij}^{ \rm  (D)}(\textbf{x},\,\chi)  \\~\\
     \bar{\sigma}_{ \rm  ij}^{ \rm  (V)}(\textbf{x},\,\chi) 	\;=	&  3\; K\; \bar{{e}}_{ \rm  ij}^{ \rm  (V)}(\textbf{x},\,\chi)  \\~\\
   \end{array}
   \right)
 %  \]
 \ea
 the symbols $^{ \rm  (D)},^{ \rm  (V)}$ marking the tensors' deviatoric and volumetric parts, respectively; $\,f\,$ denoting the reaction force given by the equation (\ref{28b});
 $\,\chi\,$ being the forcing frequency; and overbar serving to remind that we are now dealing with Fourier or Laplace images.

 To sum up, ~if the solution to an elastic problem is known in the time domain, the solution to a corresponding viscoelastic problem can be obtained in three steps: i) transform the elastic solution to the frequency domain, ii) substitute $\,\bar{\mu}(\chi)\,$ for $\,\mu\,$, iii) transform the solution back to the time domain. For a Maxwell body, we have:
 \begin{equation}
 \bar{\mu}(\chi)\;=\;\mu\;\frac{i\,\chi\,\tau}{1\,+\,i\,\chi\,\tau}~,
 \end{equation}
 while in the case of the Kelvin-Voigt model the complex shear rigidity is
 \begin{equation}
 \bar{\mu}(\chi)\;=\;\mu\;(1\,+i\,\chi\,\tau)~,
 \end{equation}
 where $\,\tau = \eta/\mu\,$.

 Applying the correspondence principle, we easily obtain for these models the deviatoric strain caused by a deviatoric stress oscillating at a frequency $\,\chi\,$, with a magnitude $\,\sigma_0\,$:\\

  For $\;\sigma^{ \rm  (D)}     =\,\sigma_0\,\cos \chi t\,$:
 \begin{center}
 \begin{tabular}{l|l}
 ${e}^{ \rm  (D,elastic)} \;= $ 	  & $\frac{\textstyle 1}{\textstyle 2\,\mu}\;\sigma_0\;\cos \chi t$ 	\\~\\
 ${e}^{ \rm  (D,Maxwell)} \;= $ 	  & $\frac{\textstyle 1}{\textstyle 2\,\mu}\;\sigma_0\;\cos \chi t +  \frac{\textstyle 1}{\textstyle 2\,\eta\, \chi}\;\sigma_0\;\sin \chi t$  \\~\\
 ${e}^{ \rm  (D,Kelvin-Voigt)} =$   & $\frac{\textstyle 1}{\textstyle 2\,\mu}\;\sigma_0\;\frac{\textstyle 1}{\textstyle 1 + \chi^2 \tau^2}\;( \cos \chi t +  \chi \tau \sin \chi t)$ \\~\\
 \hline
 \end{tabular}
 \end{center}

 For $\;\sigma^{ \rm  (D)}         = \,\sigma_0\,\sin \chi t\,$:
 \begin{center}
 \begin{tabular}{l|l}
 ${e}^{ \rm  (D,elastic)} \; = $ 	 & $\frac{\textstyle 1}{\textstyle 2\,\mu}\; \sigma_0 \;\sin \chi t$\\~\\
 ${e}^{ \rm  (D,Maxwell)} \; = $ 	   &$\frac{\textstyle 1}{\textstyle 2\,\mu}\; \sigma_0 \;\sin \chi t -  \frac{\textstyle 1}{\textstyle 2\,\eta\, \chi}\;\sigma_0\;\cos \chi t$ \\~\\
 ${e}^{ \rm  (D,Kelvin-Voigt)} =$   &$\frac{\textstyle 1}{\textstyle 2\,\mu}\; \sigma_0 \;\frac{\textstyle 1}{\textstyle 1 + \chi^2 \tau^2} ( \sin \chi t -  \chi \tau \cos \chi t)$\\~\\
 \end{tabular}
 \end{center}
 As we explained in Section \ref{4.1}, only the deviatoric parts of the tensors are affected by the presence of viscosity, the rationale for this being that a viscoelastic material is still elastic in dilatation. In other words, the bulk viscosity $\,\zeta\,$ for most solids is very high and, accordingly, the timescale $\,\tau\equiv K/\zeta\,$ is very small ($\,K\,$ being the bulk rigidity). This trivialises the computation of the volumetric part of the strain:
 \begin{equation}
 {e}^{ \rm  (V,viscoelastic)}(t)\;=\;{e}^{ \rm  (V,elastic)}(t)\;=\;\frac{\textstyle 1}{\textstyle 3\;K}\;\sigma^{ \rm  (V)}(t)~.
 \end{equation}
 The described method is used in Appendix \ref{shapes} to compute the total viscoelastic strain tensor.

 It is instructive to find the time-averaged power per unit volume, $\,p=P/V\,$, generated by a simple one-dimensional deviatoric stress at one frequency, $\;\sigma\,=\,\sigma_0\,\cos\chi t\,$. An {\it{ad hoc}} model with an empirical quality factor $\,Q\,$ gives:
 \begin{eqnarray}
 p^{ \rm  (Q-model)}\,=\;\frac{2\;\chi}{Q}\;\langle W \rangle\qquad\qquad\qquad\qquad\qquad\qquad \nonumber \\
 =\;\frac{2\;\chi}{Q} \left\langle \frac{1}{2} \;\sigma\;{e}^{ \rm  (elastic)} \right\rangle
% \nonumber \\
 =\;\frac{\chi}{4\;\mu\;Q} \;\sigma_0^2 ~,\quad
 \label{powerQ}
 \end{eqnarray}
 where  $\,\langle .\,.\,. \rangle\equiv\frac{\textstyle \chi}{\textstyle 2\,\pi}\,\int_{ \rm  t=0}^{ \rm  t=T} .\,.\,.\,dt\;$.

 For a viscoelastic model, we should use the true expression of the power involving the stress and the strain rate
 \begin{equation}
 p =  \left\langle \sigma \;\stackrel{\bf\centerdot}{{e}}^{ \rm  (viscoelastic)} \right\rangle~.
 \end{equation}
 This yields, for the Maxwell model:
 \begin{equation}
 p^{ \rm  (Maxwell)}\,=\;\frac{1}{4\,\eta}\;\sigma_0^2 ~,
 \label{powermaxwell}
 \end{equation}
 and for the Kelvin-Voigt model one:
\begin{equation}
 p^{ \rm  (Kelvin-Voigt)} = \frac{1}{4\,\mu}\;\sigma_0^2\;\frac{\chi^2\,\tau}{1\,+\,\chi^2\,\tau^2} ~,
\end{equation}
 which approaches $\,p^{ \rm  (Maxwell)}\,$ at high frequencies.

 \section{Stress and strain tensors for a viscoelastic rotator}\label{shapes}

 \subsection{Oblate shape}

 Produced by the inertial forces, the stress in the viscoelastic case is the same as in the elastic case: $\sigma_{ \rm  ij}^{ \rm  (viscoelastic)} = \sigma_{ \rm  ij}^{ \rm  (elastic)}$. Since we are considering a dynamically oblate body, and since we justifiably (see Appendix \ref{difficulties}) omitted small variations of shape, the alternating part of the stress contains only two harmonics: $\,\chi_1=\omega\,$ and $\,\chi_2=2\omega\,$.

 To write down the alternating part of the stress tensor in an elastic oblate body, we borrow time-dependent terms from the solution provided in Appendix D of \cite{sharma+2005}.\,\footnote{~Note that \cite{sharma+2005} used a different initial value for the nutation angle and also set the body to precess in a different direction. Thus any harmonic angles appearing in their stress tensor~---~which are $\omega t$ or $2 \omega t$~---~should be replaced within our convention by $\,\pi/2 - \omega t\,$ and $\,\pi - 2\omega t\,$, correspondingly.} They read as
 \ba
 \sigma_{ \rm  xx}
 &=& f_{ \rm  xx}^{ \rm  (1)} \sin \omega t - f_{ \rm  xx}^{ \rm  (2)}  \cos 2 \omega t \nonumber\\
 \sigma_{ \rm  yy}
 &=& f_{ \rm  yy}^{ \rm  (1)} \cos \omega t - f_{ \rm  yy}^{ \rm  (2)} \cos 2 \omega t \nonumber\\
 \sigma_{ \rm  zz}
 &=& - f_{ \rm  zz}^{ \rm  (1)}  \cos 2 \omega t + f_{ \rm  zz}^{ \rm  (2)} \sin 2 \omega t   \label{}\\
 \sigma_{ \rm  xy}
 &=& f_{ \rm  xy}^{ \rm  (1)}  \sin \omega t + f_{ \rm  xy}^{ \rm  (2)} \cos \omega t + f_{ \rm  xy}^{ \rm  (3)} \sin 2 \omega t   \nonumber\\
 \sigma_{ \rm  xz}
 &=& f_{ \rm  xz}^{ \rm  (1)} \sin \omega t + f_{ \rm  xz}^{ \rm  (2)} \cos \omega t - f_{ \rm  xz}^{ \rm  (3)} \cos 2 \omega t + f_{ \rm  xz}^{ \rm  (4)} \sin 2 \omega t  \nonumber\\
 \sigma_{ \rm  yz}
 &=& f_{ \rm  yz}^{ \rm  (1)} \sin \omega t + f_{ \rm  yz}^{ \rm  (2)} \cos \omega t - f_{ \rm  yz}^{ \rm  (3)} \cos 2 \omega t + f_{ \rm  yz}^{ \rm  (4)} \sin 2 \omega t \nonumber
 \ea
 with the coefficients given by
  \begin{eqnarray}
   f_{ \rm  xx}^{ \rm  (1)} =      -2  \rho  A_1  \Omega_3 \Omega_{ \rm  \perp}  y z      \;\;,\qquad  f_{ \rm  xx}^{ \rm  (2)}&=&      \rho  \Omega_{ \rm  \perp}^2 B_1  \nonumber\\
   f_{ \rm  yy}^{ \rm  (1)} =      -2  \rho  A_1  \Omega_3 \Omega_{ \rm  \perp}  x z      \;\;,\qquad  f_{ \rm  yy}^{ \rm  (2)}&=&  -   \rho  \Omega_{ \rm  \perp}^2 B_1  \nonumber\\
   f_{ \rm  zz}^{ \rm  (1)} =      A_4 \rho  h^2 \Omega_{ \rm  \perp}^2  (x^2 - y^2)      \;\;,\;\;  f_{ \rm  zz}^{ \rm  (2)}&=&  - 2 \rho x y  A_4 h^2 \Omega_{ \rm  \perp}^2  \nonumber\\
   f_{ \rm  xy}^{ \rm  (1)} =       \rho A_1 \Omega_3 \Omega_{ \rm  \perp} x z            \;\;,\;\,\quad\qquad  f_{ \rm  xy}^{ \rm  (2)}&=&  \rho A_1 \Omega_3 \Omega_{ \rm  \perp} y z   \nonumber\\
   f_{ \rm  xy}^{ \rm  (3)} =     -  \rho  \Omega_{ \rm  \perp}^2 B_1                     \;\;,\;\qquad\;\qquad\,\qquad           &~&                \qquad\qquad\qquad\qquad                     \label{}\\
   f_{ \rm  xz}^{ \rm  (1)} =      \rho A_1 h^2 \Omega_3 \Omega_{ \rm  \perp} x y         \;\;,\qquad\;\;  f_{ \rm  xz}^{ \rm  (2)}&=&   \rho \Omega_3 \Omega_{ \rm  \perp} B_4 \nonumber\\
   f_{ \rm  xz}^{ \rm  (3)} =     -  \rho  A_4 \Omega_{ \rm  \perp}^2 x z                 \;\;,\qquad\qquad  f_{ \rm  xz}^{ \rm  (4)}&=&    \rho  A_4 \Omega_{ \rm  \perp}^2 y z \nonumber\\
   f_{ \rm  yz}^{ \rm  (1)} =      \rho  \Omega_3 \Omega_{ \rm  \perp} B_4                \;\;,\qquad\,\qquad\;\; f_{ \rm  yz}^{ \rm  (2)}&=&    \rho A_1 h^2 \Omega_3 \Omega_{ \rm  \perp} x y \nonumber\\
   f_{ \rm  yz}^{ \rm  (3)} =      \rho A_4 \Omega_{ \rm  \perp}^2 y z                    \;\;,\qquad\,\qquad\;\;  f_{ \rm  yz}^{ \rm  (4)}&=&   \rho A_4 \Omega_{ \rm  \perp}^2 x z ~~~, \nonumber
  \end{eqnarray}
 where $\,h=c/a\,$ is the oblateness coefficient; $\,\rho\,$ is the density; $\;x,\,y,\,z\,$ are coordinates of a point inside the body; $\,\Omega_3\,$ and $\,\Omega_{ \rm  \perp}\,$ are defined by the equation (\ref{omegafreq}); while $\,A_1,\,A_4,\,B_1\,$ are shape coefficients introduced in \cite{sharma+2005}.
 Owing to the correspondence principle described in the previous section, the strain tensor for a Maxwell rheology is
 \begin{eqnarray}
 {e}_{ \rm  xx}^{ \rm  (Maxwell)} &=& \bigg[- \frac{f_{ \rm  yy}^{ \rm  (1)}}{6 \mu} - \frac{f_{ \rm  xx}^{ \rm  (1)}}{3 \eta \omega} + \frac{f_{ \rm  yy}^{ \rm  (1)}}{9K} \bigg] \cos \omega t
                                \nonumber\\
                              &+&
                                 \bigg[  \frac{f_{ \rm  xx}^{ \rm  (1)}}{3 \mu} - \frac{f_{ \rm  yy}^{ \rm  (1)}}{6 \eta \omega} + \frac{f_{ \rm  xx}^{ \rm  (1)}}{9K} \bigg] \sin \omega t
                                 \nonumber\\
                                 &+&
                                 \bigg[  \frac{f_{ \rm  zz}^{ \rm  (1)}}{6 \mu} + \frac{f_{ \rm  zz}^{ \rm  (2)}}{12 \eta \omega} - \frac{f_{ \rm  zz}^{ \rm  (1)}}{9K} \bigg] \cos 2 \omega t
                                 \nonumber\\
                                 &+&
                                 \bigg[  - \frac{f_{ \rm  zz}^{ \rm  (2)}}{6 \mu} + \frac{f_{ \rm  zz}^{ \rm  (1)} - 3 f_{ \rm  xx}^{ \rm  (2)}}{12 \eta \omega} + \frac{f_{ \rm  zz}^{ \rm  (2)}}{9K} \bigg] \sin 2 \omega t
 \;,\quad\quad
 \end{eqnarray}
 \begin{eqnarray}
 {e}_{ \rm  yy}^{ \rm  (Maxwell)} &=& \bigg[ \frac{f_{ \rm  yy}^{ \rm  (1)}}{3 \mu} + \frac{f_{ \rm  xx}^{ \rm  (1)}}{6 \eta \omega} + \frac{f_{ \rm  yy}^{ \rm  (1)}}{9K} \bigg] \cos \omega t
                                \nonumber\\
                              &+&
                                 \bigg[ - \frac{f_{ \rm  xx}^{ \rm  (1)}}{6 \mu} + \frac{f_{ \rm  yy}^{ \rm  (1)}}{3 \eta \omega} + \frac{f_{ \rm  xx}^{ \rm  (1)}}{9K} \bigg] \sin \omega t
                                 \nonumber\\
                                 &+&
                                 \bigg[  \frac{f_{ \rm  zz}^{ \rm  (1)} + 3 f_{ \rm  xx}^{ \rm  (2)}}{6 \mu} + \frac{f_{ \rm  zz}^{ \rm  (2)}}{12 \eta \omega} - \frac{f_{ \rm  zz}^{ \rm  (1)}}{9K} \bigg] \cos 2 \omega t
                                 \nonumber\\
                                 &+&
                                 \bigg[  - \frac{f_{ \rm  zz}^{ \rm  (2)}}{6 \mu} + \frac{f_{ \rm  zz}^{ \rm  (1)} + 3 f_{ \rm  xx}^{ \rm  (2)}}{12 \eta \omega} + \frac{f_{ \rm  zz}^{ \rm  (2)}}{9K} \bigg] \sin 2 \omega t
 \;,\quad\quad
 \end{eqnarray}
 \begin{eqnarray}
 {e}_{ \rm  zz}^{ \rm  (Maxwell)} &=& \bigg[ - \frac{f_{ \rm  yy}^{ \rm  (1)}}{6 \mu} + \frac{f_{ \rm  xx}^{ \rm  (1)}}{6 \eta \omega} + \frac{f_{ \rm  yy}^{ \rm  (1)}}{9K} \bigg] \cos \omega t
                                \nonumber\\
                              &+&
                                 \bigg[ - \frac{f_{ \rm  xx}^{ \rm  (1)}}{6 \mu} - \frac{f_{ \rm  yy}^{ \rm  (1)}}{6 \eta \omega} + \frac{f_{ \rm  xx}^{ \rm  (1)}}{9K} \bigg] \sin \omega t
                                 \nonumber\\
                                 &+&
                                 \bigg[ - \frac{f_{ \rm  zz}^{ \rm  (1)} }{3 \mu} - \frac{f_{ \rm  zz}^{ \rm  (2)}}{6 \eta \omega} - \frac{f_{ \rm  zz}^{ \rm  (1)}}{9K} \bigg] \cos 2 \omega t
                                 \nonumber\\
                                 &+&
                                 \bigg[   \frac{f_{ \rm  zz}^{ \rm  (2)}}{3 \mu} - \frac{f_{ \rm  zz}^{ \rm  (1)} }{6 \eta \omega} + \frac{f_{ \rm  zz}^{ \rm  (2)}}{9K} \bigg] \sin 2 \omega t
 \;,\quad\;\quad
 \end{eqnarray}
 \begin{eqnarray}
{e}_{ \rm  xy}^{ \rm  (Maxwell)} &=&     \bigg( - \frac{f_{ \rm  xx}^{ \rm  (1)}}{\mu}  + \frac{f_{ \rm  yy}^{ \rm  (1)}}{\eta \omega} \bigg) \cos \omega t \nonumber\\
                             &+&
                                    \bigg(  - \frac{f_{ \rm  yy}^{ \rm  (1)}}{\mu} - \frac{f_{ \rm  xx}^{ \rm  (1)}}{ \eta \omega}   \bigg) \sin \omega t \nonumber\\
                             &-&
                                   \frac{f_{ \rm  xx}^{ \rm  (2)}}{4 \eta  \omega} \cos 2 \omega t
				          - \frac{f_{ \rm  xx}^{ \rm  (2)}}{2 \mu} \sin 2 \omega t
 ~~~,\qquad\quad
 \end{eqnarray}
 \begin{eqnarray}
{e}_{ \rm  xz}^{ \rm  (Maxwell)} &=&    \bigg( \frac{f_{ \rm  xz}^{ \rm  (2)}}{2 \mu}  - \frac{f_{ \rm  xz}^{ \rm  (1)}}{2 \eta \omega} \bigg) \cos \omega t \nonumber\\
                             &+&
                                   \bigg( \frac{f_{ \rm  xz}^{ \rm  (1)} }{2 \mu} + \frac{f_{ \rm  xz}^{ \rm  (2)}}{2 \eta \omega}  \bigg) \sin \omega t  \nonumber\\
                             &+&
                                  \bigg( - \frac{f_{ \rm  xz}^{ \rm  (3)}}{2 \mu}  - \frac{f_{ \rm  xz}^{ \rm  (4)}}{4 \eta  \omega} \bigg)\cos 2 \omega t \nonumber\\
                             &+&
                                   \bigg( \frac{f_{ \rm  xz}^{ \rm  (4)}}{2 \mu} - \frac{f_{ \rm  xz}^{ \rm  (3)}}{4 \eta  \omega} \bigg) \sin 2 \omega t
  \;,\qquad\quad
 \end{eqnarray}
 \begin{eqnarray}
{e}_{ \rm  yz}^{ \rm  (Maxwell)} &=&    \bigg( \frac{f_{ \rm  xz}^{ \rm  (1)}}{2 \mu} - \frac{f_{ \rm  xz}^{ \rm  (2)}}{2 \eta \omega} \bigg) \cos \omega t \nonumber\\
                             &+&
                                   \bigg( \frac{f_{ \rm  xz}^{ \rm  (2)} }{2 \mu} +  \frac{f_{ \rm  xz}^{ \rm  (1)}}{2 \eta \omega}  \bigg) \sin \omega t \nonumber\\
                             &+&
                                  \bigg( - \frac{f_{ \rm  xz}^{ \rm  (4)}}{2 \mu}  + \frac{f_{ \rm  xz}^{ \rm  (3)}}{4 \eta  \omega} \bigg)\cos 2 \omega t \nonumber\\
                             &+&
                                   \bigg( - \frac{f_{ \rm  xz}^{ \rm  (3)}}{2 \mu}  - \frac{f_{ \rm  xz}^{ \rm  (4)}}{4 \eta  \omega} \bigg) \sin 2 \omega t
 \;.\qquad\quad
 \end{eqnarray}

 \subsection{Prism shape}

 Now consider the stress tensor for a  prism of dimensions $\,2a \times 2a \times 2c\,$, with $\,a>c\,$. Owing to dynamical oblateness and to our neglect of the small variations of shape, the alternating part of the stress contains only two harmonics: $\,\chi_1=\omega\,$ and $\,\chi_2=2\omega\,$. As was explained in the preceding subsection, the stress in the viscoelastic case coincides with that in an elastic body of the same shape: $\,\sigma_{ \rm  ij}^{ \rm  (viscoelastic)}\,=\, \sigma_{ \rm  ij}^{ \rm  (elastic)}\,$.

 The stress tensor for a prism was computed by \cite{efroimskylazarian2000}. As the authors used approximate boundary conditions and did not impose the constraints determined by the compatibility equations (see a discussion in \cite{breiter+2012}), the stress tensor must be considered as approximate.

 The stress tensor reads as
 \begin{equation} \begin{split}
 \sigma_{ \rm  xx}^{ \rm  (viscoelastic)} &=& f_{ \rm  xx} \cos 2 \omega t \\
 \sigma_{ \rm  yy}^{ \rm  (viscoelastic)} &=& f_{ \rm  yy} \cos 2 \omega t \\
 \sigma_{ \rm  zz}^{ \rm  (viscoelastic)} &=& 0 \\
 \sigma_{ \rm  xy}^{ \rm  (viscoelastic)} &=& f_{ \rm  xy} \sin 2 \omega t  \\
 \sigma_{ \rm  xz}^{ \rm  (viscoelastic)} &=& f_{ \rm  xz} \cos \omega t  \\
 \sigma_{ \rm  yz}^{ \rm  (viscoelastic)} &=& f_{ \rm  yz} \sin \omega t \\
  \end{split} \end{equation}
 with the coefficients
\begin{equation}
\begin{split}
f_{ \rm  xx} &=  \frac{\rho \Omega_{ \rm  \perp}^2}{4} (x^2 - a^2)\\
f_{ \rm  yy} &= - \frac{\rho \Omega_{ \rm  \perp}^2}{4} (y^2 - a^2)\\
f_{ \rm  xy} &=  \frac{\rho \Omega_{ \rm  \perp}^2}{4} (x^2 + y^2 - 2 a^2) \\
f_{ \rm  xz} &=  \frac{\rho \Omega_{ \rm  \perp} \Omega_3}{2} [H (z^2 - c^2) + (2 - H)(x^2 - a^2)]  \\
f_{ \rm  yz} &=  \frac{\rho \Omega_{ \rm  \perp} \Omega_3}{2} [H (z^2 - c^2) + (2 - H)(y^2 - a^2)] \\
\end{split}
\end{equation}
 where all the notations are the same as in the previous section, and $\,H=I_3/I = 2/(1+h^2)\,$. Applying the correspondence principle, we obtain the viscoelastic strain tensor for the Maxwell rheology:
\begin{equation}
\begin{split}
{e}_{ \rm  xx}^{ \rm  (Maxwell)} &= \bigg( \frac{f_{ \rm  xx}-g}{2 \mu} + \frac{g}{3K} \bigg) \cos 2 \omega t + \frac{f_{ \rm  xx}-g}{2 \eta 2 \omega} \sin 2 \omega t \\
{e}_{ \rm  yy}^{ \rm  (Maxwell)} &= \bigg( \frac{f_{ \rm  yy}-g}{2 \mu} + \frac{g}{3K} \bigg) \cos 2 \omega t + \frac{f_{ \rm  yy}-g}{2 \eta 2 \omega} \sin 2 \omega t \\
{e}_{ \rm  zz}^{ \rm  (Maxwell)} &= \bigg( \frac{-g}{2 \mu} + \frac{g}{3K} \bigg) \cos 2 \omega t + \frac{-g}{2 \eta 2 \omega} \sin 2 \omega t \\
{e}_{ \rm  xy}^{ \rm  (Maxwell)} &= - \frac{f_{ \rm  xy}}{2 \eta 2 \omega} \cos 2 \omega t + \frac{f_{ \rm  xy}}{2 \mu} \sin 2 \omega t \\
{e}_{ \rm  xz}^{ \rm  (Maxwell)} &=  \frac{f_{ \rm  xz}}{2 \mu} \cos \omega t + \frac{f_{ \rm  xz}}{2 \eta \omega} \sin \omega t \\
{e}_{ \rm  yz}^{ \rm  (Maxwell)} &= - \frac{f_{ \rm  yz}}{2 \eta \omega} \cos \omega t + \frac{f_{ \rm  yz}}{2 \mu} \sin \omega t \\
\end{split}
\end{equation}
with $\,g \equiv (f_{ \rm  xx} + f_{ \rm  yy})/3\,$.

%%%%%%%%%%%%%%%%%%%%%%%%%%%%%%%%%%%%%%%%%%%%%%%%%%

% Don't change these lines
\bsp	% typesetting comment
\label{lastpage}
\end{document}